\documentclass[a4paper,11pt]{article}
\pdfoutput=1 
\usepackage{jheppub} 
\usepackage{amsmath,bbm,graphicx,bm,slashed,multirow,color,bbold}
\usepackage{ulem}
\usepackage[T1]{fontenc}

\DeclareMathOperator\Tr{Tr}
\DeclareMathOperator\tr{tr}
\DeclareMathOperator\diag{diag}

\newcommand{\1}{\mathbb{1}}
\newcommand{\der}{\partial}

\newcommand{\RF}{R^F_k}
\newcommand{\RB}{R^B_k}
\newcommand{\DD}{\slashed{D}}
\newcommand{\dd}{\text{d}}
\newcommand{\cchi}{{\tilde\chi}}

\newcommand{\ckakko}[1]{\left\{#1\right\}}
\newcommand{\mkakko}[1]{\left({#1}\right)}
\newcommand{\kkakko}[1]{\big\langle{#1}\big\rangle}
\newcommand{\calO}{\mathcal{O}}

\newcommand{\SU}{\text{SU}}
\newcommand{\U}{\text{U}}

\renewcommand{\epsilon}{\varepsilon}
\renewcommand{\bar}[1]{\overline{#1}} 
\renewcommand{\emph}[1]{\textit{#1}}

\title{\boldmath 
Magnetic susceptibility of a strongly interacting thermal medium with $2+1$
quark flavors}

\author[a]{Kazuhiko Kamikado}
\author[b]{and Takuya Kanazawa}

\affiliation[a]{Theoretical Research Division, Nishina Center, RIKEN, Wako, 
Saitama 351-0198, Japan} \affiliation[b]{Quantum Hadron Physics
Laboratory, RIKEN, Wako, Saitama 351-0198, Japan}

\emailAdd{kazuhiko.kamikado@riken.jp}
\emailAdd{takuya.kanazawa@riken.jp}

\preprint{RIKEN-QHP-163}

\abstract{ Thermodynamics of the three-flavor quark-meson model with
$\U_{\rm A}(1)$ anomaly is studied in the presence of external magnetic
fields.  The nonperturbative functional renormalization group is
employed in order to incorporate quantum and thermal fluctuations beyond
the mean-field approximation.  We calculate the magnetic susceptibility
with proper renormalization and find that the system is diamagnetic in
the hadron phase and paramagnetic in the hot plasma phase.  The obtained
values of the magnetic susceptibility are in reasonable agreement with
the data from first-principle lattice QCD.  Comparison with the
mean-field approximation, the Hadron Resonance Gas model and a free gas
with temperature-dependent masses is also made.  }
 
\begin{document}
 \maketitle
 \flushbottom

 \section{Introduction}
 
 Recent years have seen a surge of interest in QCD in external magnetic fields. 
 The electromagnetic response of QCD is not only interesting as a theoretical
 probe to the dynamics of QCD, but also important in experimental heavy 
 ion collisions, cosmology and astrophysics. Special neutron stars called magnetars 
 possess a strong surface magnetic field reaching $10^{10}$ T \cite{Duncan:1992hi,Harding:2006qn} whereas 
 the primordial magnetic field in early Universe is estimated to be even 
 as large as $\sim 10^{19}$ T \cite{Grasso:2000wj}. In non-central heavy
 ion collisions at RHIC and LHC, an instantaneous magnetic field of strength $\sim
 10^{15}$ T perpendicular to the reaction plane could be produced and
 may have impact on the thermodynamics of the quark-gluon plasma, which could 
 lead to experimentally useful signatures \cite{Kharzeev:2007jp,Fukushima:2008xe,Skokov:2009qp}.
 
 The effect of magnetic field on the chiral dynamics of QCD at finite
 temperature $T>0$ has been vigorously investigated in chiral effective
 models (see \cite{Gatto:2012sp,Shovkovy:2012zn} for reviews).  Among
 other things it was found that the magnetic field catalyzes spontaneous
 chiral symmetry breaking at least for low $T$, an effect called
 \emph{magnetic catalysis}. This model-independent phenomenon is
 explained through dimensional reduction ($3+1 \to 1+1$) in the quark
 sector in a magnetic field \cite{Gusynin:1994xp,Gusynin:1995nb}. The
 dynamics of QCD in a magnetic field has also been studied in lattice
 simulations
 \cite{Buividovich:2008wf,D'Elia:2010nq,D'Elia:2011zu,Braguta:2011hq,Bali:2011qj,Ilgenfritz:2012fw,
 Luschevskaya:2012xd,Bali:2012zg,Bali:2012jv,Bali:2013esa,Bonati:2013lca,Levkova:2013qda,
 Ilgenfritz:2013ara,Bonati:2013vba,Bornyakov:2013eya}, see
 \cite{D'Elia:2012tr} for a review. Simulations at the physical quark
 mass \cite{Bali:2011qj,Bali:2012zg} showed that the effect of a
 magnetic field is non-monotonic: although the chiral condensate
 increases at low temperature, it \emph{decreases} at high temperature,
 resulting in a lower chiral pseudo-critical temperature $T_c$ in a
 stronger magnetic field. Gluonic observables also show a similar
 behavior \cite{Bali:2013esa}.  Various attempts have been made to
 explain the origin of this \emph{inverse magnetic catalysis}
 \footnote{There is a similar phenomenon called by the same name but is
 specific to nonzero chemical potential \cite{Preis:2010cq}. This seems
 to have a different origin from that of inverse magnetic catalysis at
 zero density; the interested reader is referred to a review
 \cite{Preis:2012fh}.}  (or \emph{magnetic inhibition})
 \cite{Agasian:2008tb,Fraga:2012fs,
 Fukushima:2012kc,Bruckmann:2013oba,Chao:2013qpa,Fraga:2013ova,
 Kamikado:2013pya,Farias:2014eca,Ferreira:2014kpa,
 Ayala:2014iba,Ayala:2014gwa,Ferrer:2014qka,Fayazbakhsh:2014mca}.
 
 Lattice measurements of quantities such as the pressure, energy density, 
 entropy density and magnetization have revealed thermodynamic properties of QCD 
 in external magnetic fields. In particular, the \emph{magnetic susceptibility}%
 \footnote{%
   To avoid confusion, we remark that a quantity called \emph{magnetic
   susceptibility of the quark condensate} is defined with regard to the tensor polarization
   $\langle\bar\psi\sigma_{\mu\nu}\psi\rangle$ \cite{Ioffe:1983ju} and 
   has been measured on the lattice \cite{Buividovich:2009ih,Bali:2012jv} 
   (for theoretical works, see \cite{Frasca:2011zn} and references therein). 
   However, this ``susceptibility'' is just the spin-related piece of the full magnetic susceptibility
   defined from the pressure in a magnetic field. In this paper we
   focus on the latter \emph{full} magnetic susceptibility, not the former.
 }   
 $\chi(T)$, which quantifies the leading response of pressure to weak external magnetic fields, 
 has been accurately measured in recent lattice simulations with light dynamical quarks 
 \cite{Bonati:2013lca,Levkova:2013qda,Bonati:2013vba,Bali:2013owa,Bali:2014kia}, 
 where various methods were adopted to circumvent the flux quantization condition 
 on a periodic lattice, as summarized in \cite[Sec.~3.1]{Bali:2014kia}. 
 The simulations revealed that QCD at $T\gtrsim T_c$ is strongly \emph{paramagnetic}, 
 characterized by $\chi>0$, whereas the behavior at $T\lesssim 100$ MeV is 
 consistent with \emph{weak diamagnetism} with $\chi\lesssim 0$. Thus, 
 interestingly, QCD seems to undergo a characteristic change in its magnetic profile 
 at finite temperature, despite the absence of a genuine phase transition.  
 On the theoretical side, the magnetization and magnetic susceptibility 
 of QCD at zero density have been studied in, e.g., perturbative QCD \cite{Cohen:2008bk}, 
 the Hadron Resonance Gas model \cite{Endrodi:2013cs}, 
 holographic QCD \cite{Bergman:2008sg}, a transport model \cite{Steinert:2013fza}, 
 a potential model \cite{Kabat:2002er}, a non-interacting quark gas 
 with the Polyakov loop \cite{Orlovsky:2014cta}, 
 spatially compactified QCD \cite{Anber:2013tra} and  
 the $\SU(3)$ linear sigma model with the Polyakov loop \cite{Tawfik:2014hwa}. 

 In this work, we apply the functional renormalization group (FRG)
 \cite{Wetterich:1992yh} to the quark-meson (QM) model with three quark
 flavors to study magnetic properties of QCD at finite temperature and zero density. 
 FRG is a powerful
 nonperturbative method to go beyond the mean-field approximation by
 fully taking thermal and quantum fluctuations into account; see
 \cite{Berges:2000ew,Pawlowski:2005xe,Delamotte:2007pf,Braun:2011pp} for
 reviews. While FRG has already been applied to chiral models in a
 magnetic field
 \cite{Skokov:2011ib,Scherer:2012nn,Fukushima:2012xw,Andersen:2012bq,
 Andersen:2013swa,Kamikado:2013pya}, so far no attempt has been made to
 include strangeness. Here we shall extend the FRG analysis of 
 the $N_f=3$ QM model \cite{Mitter:2013fxa} to the case of 
 nonzero magnetic fields and study its thermodynamic behavior, with particular 
 emphasis on the magnetic susceptibility, $\chi(T)$. 
 
 While it seems well recognized by now that chiral effective models do not support the inverse magnetic 
 catalysis of QCD in a \emph{strong} magnetic field, it does not necessarily imply that they also fail 
 to explain the behavior of QCD in a \emph{weak} magnetic field. 
 Refs.~\cite{Farias:2014eca,Ferreira:2014kpa,Ayala:2014iba,Ferrer:2014qka} suggest that 
 the running of the QCD coupling with a magnetic field could be the origin of the inverse magnetic catalysis,   
 but such a factor is not likely to play a dominant role in QCD in an infinitesimal magnetic field. 
 This line of reasoning leaves open the possibility that chiral models may be able to capture physics 
 at weak external fields correctly. Our aim is to put this expectation to the test 
 through a stringent quantitative comparison between lattice data and the FRG calculation. 

 This paper is organized as follows.  
 In section \ref{sec:formulation}, we introduce the $N_f=3$ QM model and describe the
 formulation of FRG. We specify our truncation of the scale-dependent quantum 
 effective action, introduce regulator functions and present the flow equation. 
 In section \ref{sec:numerical-results}, we show plots of physical observables
 obtained by solving the flow equation numerically and discuss their characteristics.  
 The main results are summarized in Figures \ref{fig:sus_vs_lat_normal_t} and 
  \ref{fig:sus_vs_lat_shifted_t}. 
 We compare the magnetic susceptibility from $N_f=3$ FRG with results from recent 
 lattice simulations, $N_f=2$ FRG, the mean-field calculation, and non-interacting quark-meson gas 
 with temperature-dependent masses.  
 Section \ref{sec:conclusion} is devoted to conclusions. Technical details of the flow 
 equation and the derivation of pressure of a non-interacting gas are relegated 
 to the appendices. 
 
 In this paper we will use natural units and the Heaviside-Lorentz conventions, 
 in which $\epsilon_0=\mu_0=1$ and the fine-structure constant $\alpha_{\rm em}=e^2/4\pi\approx 
 1/137$. The magnitude of a magnetic field $\bm{B}$ will be measured in the combination 
 $eB$, which is useful in natural units.

 \section{Formulation}
 \label{sec:formulation} 
 
 In Section \ref{sec:Nf3QM} we introduce the three-flavor QM model as an effective 
 model with the same chiral symmetry breaking pattern as in QCD. In Section \ref{sec:frge} 
 we review basics of FRG and explain how mesonic fluctuations which are neglected
 in the mean-field approximation are incorporated by the non-perturbative 
 flow equation. Finally in Section \ref{sec:obs}, we explain how to compute observables 
 from the effective potential of the model.

  \subsection[$N_f=3$ Quark-Meson model]{\boldmath $N_f=3$ Quark-Meson model}
  \label{sec:Nf3QM}
  
  The $N_f=3$ QM model \cite{Jungnickel:1995fp,Schaefer:2008hk,
  Schaefer:2009ui,Chatterjee:2011jd,Schaefer:2011ex,Mintz:2012mz,
  Stiele:2013pma,Mitter:2013fxa,Beisitzer:2014kea,Tawfik:2014hwa} is 
  the $\U(3)$ linear sigma model
  \cite{Levy:1967,Gasiorowicz:1969kn,Schechter:1971qa,Schechter:1993tc,Chan:1974rb,Napsuciale:1998ip,Lenaghan:2000ey,
  Roder:2003uz,Parganlija:2012fy} coupled to quarks:
  \begin{equation}
   \begin{split}
    {\cal L} & = 
    \sum_{i=1}^{N_c}
    \bar{\psi}_i \Big[\DD + g \sum_{a=0}^{8} T_a(\sigma_a + i \gamma_5
    \pi_a)\Big] \psi_i +
    \tr \!\big[{\cal D}_{\mu} \Sigma \,({\cal D}_{\mu} \Sigma)^{\dagger} \big]
    \\
    & \qquad + U(\rho_{1},\rho_{2}) - h_x \sigma_x - h_y \sigma_y  - c_A \xi \,. 
   \end{split}
  \end{equation}
  Several remarks are in order.
  \begin{itemize}
    \item 
    $\psi=(u,d,s)^T$ represents the three-flavor quark field with three colors $(N_c=3)$. 
    \item 
    $\Sigma$ is a matrix field consisting of 
    the scalar ($\sigma_{a}$) and pseudo-scalar ($\pi_{a}$) meson \mbox{multiplets,}
    \begin{equation}
      \Sigma = \sum_{a=0}^{8} T_{a} (\sigma_{a} + i \pi_{a}) \,, 
    \end{equation}
    where $\{T_{a}\}$ are the nine generators of $\U(3)$ normalized
    as $ {\rm tr} [T_{a}T_{b}] = \frac{1}{2}\delta_{ab}$. They are related to 
    the Gell-Mann matrices as $T_a=\lambda_a/2$ ($a=1,\dots,8$), and $T_0=\1/\sqrt{6}$. 
    \item 
    The covariant derivatives are defined as 
    \begin{align}
      \DD \equiv \gamma_\mu(\partial_\mu - iQA_\mu)
      \qquad \text{and}\qquad 
      \mathcal{D}_\mu \Sigma \equiv \partial_\mu \Sigma - i A_\mu \big[ Q,\Sigma \big]\,, 
    \end{align}
    with 
    $\displaystyle Q = \diag\bigg(\frac{2}{3}e,\ -\frac{1}{3}e,\ -\frac{1}{3}e \bigg)$,  
    $\vec{A} =(0,Bx_1,0)$ and $A_4=0$ \,. 
    \item $U(\rho_1,\rho_2)$ is the $\U(3)\times \U(3)$-invariant part
	  of the
    bosonic potential written in terms of two invariants%
    \footnote{To avoid confusion, we note that our $\rho_2$ defined in \eqref{eq:r1r2} 
    coincides with $\tilde\rho_2$ in \cite{Mitter:2013fxa} .}  
    \begin{equation}
        \label{eq:r1r2}
        \rho_{1} = {\rm tr} [\Sigma \Sigma^{\dagger}]  \quad \text{and}\quad 
        \rho_2 =\tr\!\Big[ \Big( \Sigma \Sigma^{\dagger} - \frac{1}{3}\rho_1 \1\Big)^2 \Big]\,.
    \end{equation}
    By definition, $\rho_1\geq 0$ and $\rho_2\geq 0$. 
    In this work we neglect the possible dependence of $U$ 
    on another invariant ${\rm tr}[(\Sigma \Sigma^{\dagger})^3]$ 
    for simplicity.%
    \footnote{Generally, we have $N$ independent invariants for $\U(N)
    \times \U(N)$ chiral flavor rotation \cite{Jungnickel:1995fp}.} 
    \item 
    The linear terms $h_x\sigma_x+h_y\sigma_y$ stand for 
    the explicit chiral symmetry breaking due to nonzero quark masses.  
    The strange-nonstrange basis is defined as
    \begin{align}
     \begin{pmatrix}
        \sigma_x\\
        \sigma_y
     \end{pmatrix}
     \equiv  \frac{1}{\sqrt{3}} 
     \begin{pmatrix}
        \sqrt{2} &1\\
        1 &-\sqrt{2} \\
     \end{pmatrix}
     \begin{pmatrix}
        \sigma_0 \\
        \sigma_8
     \end{pmatrix}\,.
     \label{eq:sigmaxy}
    \end{align}
    This parametrization automatically respects the $\SU(2)_{\rm V}$ symmetry of light quarks. 
    \item  
    $\xi \equiv \det \Sigma  + \det \Sigma^{\dagger}$ represents 
    the effect of the $\U_{\rm A}(1)$ anomaly. Another invariant 
    $i(\det \Sigma - \det \Sigma^\dagger)$ breaks CP invariance of the model 
    and is omitted. 
  \end{itemize}

  \subsection{Functional renormalization group equation}
  \label{sec:frge}
  
  The basic idea of FRG is to start from a tree-level classical action $\Gamma_{k=\Lambda}$ 
  at the UV scale $\Lambda$, and keep track of the flow of the scale-dependent
  effective action $\Gamma_k$ while integrating out degrees of freedom with
  intermediate momenta successively; finally at $k=0$ the full quantum
  effective action $\Gamma_{k=0}$ is obtained. 
  The functional renormalization group equation~\cite{Wetterich:1992yh}
  (called the Wetterich equation) reads
  \begin{align}
   \label{eq:FRG-}
   \der_k \Gamma_k = 
   {\frac{1}{2}\Tr\left[ \frac{1}{\Gamma^{(2,0)}_k+\RB} \der_k \RB \right]}
   - {\Tr\left[ \frac{1}{\Gamma^{(0,2)}_k+\RF} \der_k \RF \right]}\,, 
  \end{align}
  where $\RB$ and $\RF$ are cutoff functions (regulators) for bosons and fermions, 
  while $\Gamma_k^{(2,0)}$ and $\Gamma_k^{(0,2)}$ represent the second
  functional derivative of $\Gamma_k$ with respect to bosonic and fermionic fields, 
  respectively.  $\Tr$ is a trace in the functional space. 
  Although \eqref{eq:FRG-} has a simple one-loop structure, 
  \eqref{eq:FRG-} incorporates effects of arbitrarily high order
  diagrams in the perturbative expansion through the full
  field-dependent propagator \mbox{$(\Gamma^{(2)}_k+R_k)^{-1}$}. 
  Further details of FRG can be found in reviews
  \cite{Berges:2000ew,Pawlowski:2005xe,Delamotte:2007pf,Braun:2011pp}.
          
  The flow of $\Gamma_k$ from UV to IR is governed by the cutoff
  functions $R_k^{B,F}(p)$.  The latter must satisfy (i) $\displaystyle
  \lim_{k\to\infty}R_k(p)=\infty$, (ii) $\displaystyle \lim_{k\to
  0}R_k(p)=0$, and (iii) $\displaystyle \lim_{p\to 0}R_k(p)>0$
  \cite{Berges:2000ew}.  In this work we use the following regulators 
  \begin{subequations}
  \begin{align}
   \label{eq:RB_3d}
   \RB(p) & = (k^2-\vec{p}\,^2)  \,\Theta(k^2-\vec{p}\,^2) \,,
   \\
   \RF(p) & 
   = - i \,\slashed{\vec{p}}\,r_k(\vec{p}\,)\qquad \text{with }\quad 
   r_k(\vec{p}\,) \equiv \left(\frac{k}{|\vec{p}\,|}-1\right) \Theta(k^2-\vec{p}\,^2) \,,
   \label{eq:RF_3}
   \end{align}
  \end{subequations}
  for bosons and fermions, respectively. These are the finite-temperature 
  version of the so-called optimised regulator \cite{Litim:2001up}. 
  They can be used in a magnetic field as well \cite{Andersen:2012bq}. 
  With this choice of regulators, we can perform the Matsubara sum analytically.

  Although the Wetterich equation \eqref{eq:FRG-}
  formulated in the infinite-dimensional functional space is
  \emph{exact}, for practical calculations we must find a proper truncation of
  $\Gamma_k$. In this work we employ the so-called local-potential approximation (LPA), 
  which neglects anomalous dimensions of fields altogether but is commonly used due to 
  its technical simplicity. More explicitly, we use 
  the truncated effective action at the scale $k$ of the form
  \begin{equation}
   \begin{split}
    \Gamma_k [\psi, \sigma, \pi] = & \int_0^{1/T} \!\!\! dx_4 \int d^3x ~
    \Bigg \{ 
    \sum_{i=1}^{N_c} \bar\psi_i \Big[ 
      \DD + g \sum_{a=0}^{8} T_a(\sigma_a + i\gamma_5 \pi_a) 
    \Big] \psi_i 
    \\
    & \quad 
    + \tr\big[ (\mathcal{D}_\mu \Sigma)(\mathcal{D}_\mu \Sigma)^\dagger \big] +
    U_k(\rho_1,\rho_2) - h_x \sigma_x - h_y\sigma_y  - c_A \xi \Bigg\}\,. 
   \end{split}
   \label{eq:lpa_effective_action}
  \end{equation}
  Here, for a technical reason, 
  we have ignored isospin symmetry breaking in the bosonic potential 
  due to the magnetic field (see also \cite{Andersen:2013swa} for a discussion 
  on this point). 
  In the limit $\bm{B}\to 0$, \eqref{eq:lpa_effective_action} 
  reduces to the effective action used in \cite{Mitter:2013fxa}. 
  
  By using the cutoff functions (\ref{eq:RB_3d}), (\ref{eq:RF_3}), and
  the LPA effective action (\ref{eq:lpa_effective_action}), we can
  easily derive the flow equation for the symmetric part of the bosonic 
  potential \cite{Skokov:2011ib,Andersen:2012bq,Andersen:2013swa}: 
  \begin{equation}
   \begin{split}
    \partial_k U_k & =\frac{k^4}{12 \pi^2}\Bigg\{
    \sum_{b}^{} \alpha_{b}(k)
    \frac{\coth \frac{E_b(k)}{2T} }{E_b(k)}
    - \sum_{f=u,d,s} \alpha_{f}(k)
    \frac{\tanh \frac{E_f(k)}{2T} }{E_f(k)}
    \Bigg\} \,,
   \end{split}
   \label{eq:flow_equation_for_potential}
  \end{equation}
  where the index $b$ runs over all $9+9=18$ mesons. 
  $E_b(k)$ and $E_f(k)$ are scale- and field-dependent energies of mesons and fermions, 
  respectively, defined as \mbox{$E_i(k)\equiv \sqrt{k^2+M_i^2}$}. For quarks, we have
  \begin{align}
    M_{u,d} = \frac{g\sigma_x}{2} \qquad \text{and} \qquad M_s = \frac{g\sigma_y}{\sqrt{2}} \,.
    \label{eq:Mquarks}
  \end{align}
  The meson masses are summarized in appendix~\ref{sec:mass-eigenv-three}. 
  
  The dimensionless factors $\alpha_{b}(k)$ and $\alpha_{f}(k)$ are defined as
  \begin{subequations}
   \begin{align}
     \alpha_{b}(k) &= 3\frac{|e_b B|}{k^2}\sum_{n=0}^{\infty}  
     \sqrt{1 - (2n+1) \frac{|e_b B|}{k^2} } \;\Theta\left(1 - 
     (2n+1) \frac{|e_b B|}{k^2} \right)\,,
     \\
     \alpha_{f}(k) &= 6 N_c \frac{|e_f B|}{k^2} \left\{ 1 + 2\sum_{n=1}^{\infty}  
     \sqrt{1 - 2n \frac{|e_f B|}{k^2} } \;\Theta \left(1 - 2n
     \frac{|e_f B|}{k^2} \right)\right\} \,,
   \end{align} 
  \end{subequations}
  where $e_b$ and $e_f$ are the electric charge of each field.  
  For neutral particles ($e_b,\,e_f=0$), or for charged particles 
  in the limit of a vanishing magnetic field ($eB\to 0$), these factors simply become  
  $\alpha_b(k)=1$ and $\alpha_f(k)=4N_c$, which recovers the conventional 
  flow equation in the absence of  magnetic fields correctly \cite{Andersen:2012bq}.

  \subsection{Observables} 
  \label{sec:obs}
  
  Let us explain how to obtain physical quantities from the $k \rightarrow 0$ 
  limit of the flow equation. First of all, the condensates are determined from the 
  minimum of the total effective potential at $k=0$. We denote 
  the expectation values of $\sigma_x$ and $\sigma_y$ thus obtained 
  as $\langle\sigma_x\rangle_{k=0}$ and $\langle \sigma_y\rangle_{k=0}$\,, respectively. 
  The physical constituent quark masses are obtained from \eqref{eq:Mquarks} as
  \begin{align}
    M_{u,d} = \frac{g}{2} \langle\sigma_x\rangle_{k=0} 
    \quad\text{and}\quad M_s = \frac{g}{\sqrt{2}} \langle\sigma_y\rangle_{k=0} \,.
  \end{align}
  The meson screening masses can be obtained from the formulas 
  in appendix~\ref{sec:mass-eigenv-three}.
  
  The PCAC relations determine the pion and kaon decay constants
  $f_{\pi},f_{K}$ from the values of the condensates \cite{Lenaghan:2000ey,Schaefer:2008hk} as 
  \begin{equation}
     f_{\pi} = \langle \sigma_x \rangle_{k=0} 
     \quad \text{and} \quad 
     f_{K} = \frac{1}{2}
     \langle \sigma_x \rangle_{k=0} + 
     \frac{1}{\sqrt{2}} \langle \sigma_y \rangle_{k=0}\,.  
     \label{eq:decay_constant}
  \end{equation}

  The pressure follows from the minimum of the total effective 
  potential at $k=0$:  
  \begin{align}
    P_0 & = - \min_{\rho_1,\,\rho_2} \big[ U_{k=0}(\rho_1,\rho_2) 
    - h_x \sigma_x - h_y \sigma_y - c_A \xi \big] \,.
  \end{align}
  The contribution from the pure magnetic field, $B^2/2$, is left out 
  from our definition of the pressure, because it is independent of temperature 
  and does not influence observables.
  
  In addition, in the FRG method we 
  usually include a residual term ($P_r$) in the total pressure $P$ to ameliorate the 
  ultraviolet cutoff artifacts \cite{Braun:2003ii,Braun:2009si,Herbst:2010rf},
  \begin{align}
    P &= P_0 - P_0\big|_{T=B=0} + P_r  \,,
    \\ 
    P_r & \equiv - N_c \sum_{f=u,d,s}
    \sum_{s=\pm\frac{1}{2}} \sum_{n=0}^{\infty}\frac{|e_f B|}{2 \pi^2 }
    \times
    \notag
    \\
    & \quad \quad \times 
    \int_{\Lambda}^{\infty} dk~ \Theta \big(k^2-p_{\perp}^2[e_f,s,n]\big)
    \sqrt{k^2-p_{\perp}^2[e_f,s,n]} \left({\rm tanh}\frac{k}{2T} -1\right)\,,
    \label{eq:P_residue}
  \end{align}
  with $p^2_{\perp}[e_f,s,n] \equiv (2n+1-2s)|e_fB|$. The residual part 
  compensates for the pressure from modes with momentum larger than $\Lambda$ 
  and ensures that the Stephan-Boltzmann limit of free quark gas is reached 
  at sufficiently high temperatures. 
  
  From the pressure, one can extract the \emph{bare} magnetic
  susceptibility $\cchi$ as the leading response of the system to the
  external magnetic field:
  \begin{equation}
   P(T,B) = P(T,0) + \frac{\cchi(T)}{2} (eB)^2 + \calO\mkakko{B^4} \,.
    \label{eq:Pressure}
  \end{equation}
  However, the pressure contains a $B$-dependent divergence and incidentally 
  $\cchi(T)$ is also divergent. This is related to the issue of electric charge renormalization 
  in QED \cite{Endrodi:2013cs,Bali:2013owa,Bali:2014kia}. In this work, 
  we renormalize $\cchi$ by subtracting the divergent contribution at $T=0$ as
  \begin{align}
    \chi(T) & \equiv \cchi(T) - \cchi(0) \,. 
    \label{eq:definition_sus}
  \end{align}
  Thus $\chi(T=0)$ vanishes by definition. This means that the matter
  contribution to the pressure at $T=0$ begins at
  $\calO\mkakko{B^4}$. An intriguing consequence of renormalization is
  that the magnetic susceptibility obtained this way is intertwined with
  the nonperturbative IR physics at $T=0$ even at arbitrarily high
  temperatures; this phenomenon will be demonstrated explicitly for a
  non-interacting gas with temperature-dependent masses in appendix
  \ref{sc:vac_contr}.  We remark that the renormalization prescription
  \eqref{eq:definition_sus} agrees with those in recent lattice
  simulations
  \cite{Levkova:2013qda,Bonati:2013lca,Bali:2013owa,Bonati:2013vba,Bali:2014kia},
  hence allowing for a direct comparison between the present model
  calculation and the lattice data.

  In actual numerics we proceed by first evaluating the subtracted
  pressure \cite{Bonati:2013lca,Bonati:2013vba}
  \begin{equation}
    \Delta P \equiv \left(P(T,B) - P(T,0) \right) - \left( P(0,B)-P(0,0)
  \right) \,, \label{eq:normalized_pressure}
  \end{equation}
  and then measuring the magnetic susceptibility $\chi(T)$ through a
  polynomial fitting to $\Delta P$.

  There is one caveat here. We have neglected the back reaction from the
  matter, i.e., neglected the difference between the net magnetic field ($B$) and the magnetic field 
  generated by external currents only ($H$). Their relationship is given by $H=B-M$, with $M$ 
  the magnetization. For weak fields, $M$ is linear in $H$ and 
  one can define the magnetic susceptibility $\chi_{\rm back}$ 
  which incorporates the matter back-reaction by $M=e^2 \chi_{\rm back}H$. The matter 
  contribution to the pressure becomes 
  $\Delta P=\frac{1}{2}M\cdot B=\frac{1}{2}\frac{\chi_{\rm back}}{1+e^2\chi_{\rm back}}(eB)^2$. 
  Comparison with \eqref{eq:Pressure} and \eqref{eq:definition_sus} yields 
  \begin{equation}
    \chi_{\rm back} = \frac{\chi}{1-e^2\chi}\simeq \chi + e^2\chi^2 + \calO(\chi^3)\,.
  \end{equation} 
  However, as shown in section \ref{sec:numerical-results}, $\chi$ is of order $10^{-2}$ and 
  the correction from the back-reaction of matter is negligible. 
  
  Another important quantity that characterizes magnetic properties of
  the system is the magnetic permeability, $\mu$, which is related to
  the magnetic susceptibility as\footnote{In SI units, this
  corresponds to the ratio $\mu/\mu_0$.}
  \begin{align}
    \mu(T) & = \frac{1}{1-4\pi \alpha_{\rm em} \chi(T)}\,. 
  \end{align}
  Since it does not provide any new information compared to $\chi(T)$ itself, 
  we will not show a separate plot for $\mu(T)$.

 \section{Numerical implementation}
 \label{sec:numerical-results}
  \subsection{Setup}
  
  We solved the flow equation \eqref{eq:flow_equation_for_potential}
  with the two-dimensional Taylor method. Namely, we expand the
  scale-dependent effective potential around its running minimum and
  then cast \eqref{eq:flow_equation_for_potential} into a set of coupled
  flow equations for the coefficients of the expansion. All technical
  details of the Taylor method in the QM model are given in
  appendix~\ref{sec:taylor-method}.
  
  The flow equations for Taylor coefficients are then solved by
  integration from $k=\Lambda$ to $k=0$ with the Euler method, keeping
  the step size of $k$ smaller than $0.5$ MeV. We confirmed numerical
  stability of results by changing the step size.
  
  For the initial condition of the flow, we used
  \begin{equation}
    U_{k=\Lambda}(\rho_1,\rho_2) = a_\Lambda^{(1,0)} \rho_1 +
    \frac{a_\Lambda^{(2,0)}}{2}\rho_1^2 + a_\Lambda^{(0,1)} \rho_2\,,
  \end{equation}
  where $a_\Lambda^{(1,0)}, a_\Lambda^{(2,0)}$ and $a_\Lambda^{(0,1)}$ are 
  free parameters of the model.

  \begin{table*}[t]
   \centering
   \begin{tabular}{|c||c c c c c c c|c c|}
    \hline 
    & $g$
    & $a_\Lambda^{(1,0)^{\mathstrut}}/\Lambda^2$ 
    & $a_\Lambda^{(2,0)}$ 
    & $a_\Lambda^{(0,1)}$ 
    & $h_{x}/\Lambda^3$ 
    & $h_{y}/\Lambda^3$ 
    & $c_A/\Lambda$ 
    & $f_{\pi}$
    & $f_{K}$
    \\\hline \hline 
    FRG &$6.5$& $0.56$ & $20.0$ & $10.0$ & $1.76 \times 10^{-3}$ &
			    $3.79 \times 10^{-2}$ & $4.8$ &$91.8$ & $112.3$ 
    \\
    MF  & $6.5$ & $1.07$ &$5.0$ & $2.0$ & $1.76 \times 10^{-3}$ & $3.79
    			    \times 10^{-2}$ & $4.8$&$91.5$ & $113.4$
    \\\hline 
   \end{tabular}
   \caption{Initial conditions at $k=\Lambda$ and the resulting $f_{\pi}$ and $f_{K}$ 
   (in units of MeV) at $eB=T=0$ for $N_f=3$ FRG (first row) and the
   mean-field calculation (MF, second row). We used \mbox{$\Lambda=1$ GeV} in both
   calculations. }  \label{tab:paraemter_3flavor}
  \end{table*}
  
  Besides the UV cutoff scale $\Lambda$, the $N_f=3$ QM model still has
  seven free parameters: $g, a_\Lambda^{(1,0)}, a_\Lambda^{(2,0)},
  a_\Lambda^{(0,1)}, h_x, h_y$ and $c_A$. We adjusted these parameters
  so as to reproduce the pion and kaon decay constants, the light quark
  mass, and the pion/kaon/sigma/eta masses at $eB=T=0$.  We summarize
  our parameter set in Table~\ref{tab:paraemter_3flavor} and the
  resulting particle masses in Table~\ref{tab:particle_data}.
    
  \begin{table}[t]
   \begin{center}
    \begin{tabular}{|c|c|c|c||c|c|c|c|}
     \hline 
     particle & mass (MeV) &  $|q|/e$ & spin
     &particle & mass (MeV) & $|q|/e$& spin \\
     \hline\hline 
     $u$ & 298.1 & 2/3 &1/2 & 
     $s$ & 430.8 & 1/3&1/2 \\
     $d$ & 298.1 & 1/3 &1/2 & --- 
     & ---  & --- & --- \\
     \hline 
     $\pi^{0}$ & 138.4 &  0 &0&$a_0^0$ & 1028.9  & 0&0\\
     $\pi^{\pm}$ & 138.4  & $$1 &0& $a_0^{\pm}$ & 1028.9 & $$1&0\\
     $K^{0}$,$\bar{K}^0$  & 496.7 & 0&0 &$\kappa^0$, $\bar{\kappa}^0$   & 1126.8 & 0&0\\
     $K^{\pm}$  & 496.7 &  1 &0 &$\kappa^{\pm}$  & 1126.8  & $$1&0\\
     $\eta$   & 539.2  & 0 &0&$\sigma$   & 533.7  & 0&0 \\
     $\eta'$  & 959.8  & 0 &0&$f_0$   & 1237.8 & 0&0 \\
     \hline 
    \end{tabular}
    \caption{Table of quarks and mesons in the $N_f=3$ QM model. 
    Their masses are obtained from FRG at $eB=T=0$ with the parameter set in
    Table~\ref{tab:paraemter_3flavor}.
    }  
    \label{tab:particle_data}
   \end{center}
  \end{table}
  In order to evaluate the effect of mesonic fluctuations, we also performed 
  calculations in the mean-field approximation. This is simply done by 
  neglecting the first term in \eqref{eq:flow_equation_for_potential} and 
  solving the resulting flow equation. However the initial conditions
  have to be readjusted to realize the same physical observables at $k=0$; see the second row in 
  Table~\ref{tab:paraemter_3flavor}. 
  
  In addition, we also performed FRG for the two-flavor QM model 
  for the purpose of comparison with the three-flavor QM model. All details for the truncated effective action, 
  the flow equation and the initial conditions are summarized in appendix \ref{sec:two-flavor-qm}. 
  To solve the flow equation, we again used the Taylor expansion method (see 
  appendix \ref{sec:taylor-method}). The results from the $N_f=2$ FRG, the $N_f=3$ FRG and 
  the mean-field approximation ($N_f=3$) will be juxtaposed in section \ref{sec:chi}. 

  We end this subsection with a cautionary remark.  The Taylor method is
  ineffective in the case of a first-order phase transition, because a
  smooth flow of the scale-dependent minimum of the potential, on which
  the Taylor expansion is based, breaks down.  In the chiral limit, the
  phase transition in $\U(N) \times \U(N)$-symmetric models is first
  order according to the one-loop $\epsilon$-expansion
  \cite{Pisarski:1983ms}, which has been confirmed by FRG with the Grid
  method
  \cite{Berges:1996ja,Berges:1996ib,Fukushima:2010ji,Fejos:2014qga}.
  Then the applicability of the Taylor method is questionable.  However,
  as shown in \cite{Mitter:2013fxa}, for physical values of the quark
  masses and anomaly strength, the three-flavor chiral transition
  becomes a crossover at least for $\bm{B}=0$. Assuming that this is the
  case also for small nonzero magnetic fields, we can justify the usage
  of the Taylor method.

  \subsection{Results at nonzero magnetic field}
  \subsubsection{Pressure, masses and decay constants}
   
   \begin{figure}[t!]
   \begin{center}
     \includegraphics[width=0.48\columnwidth]{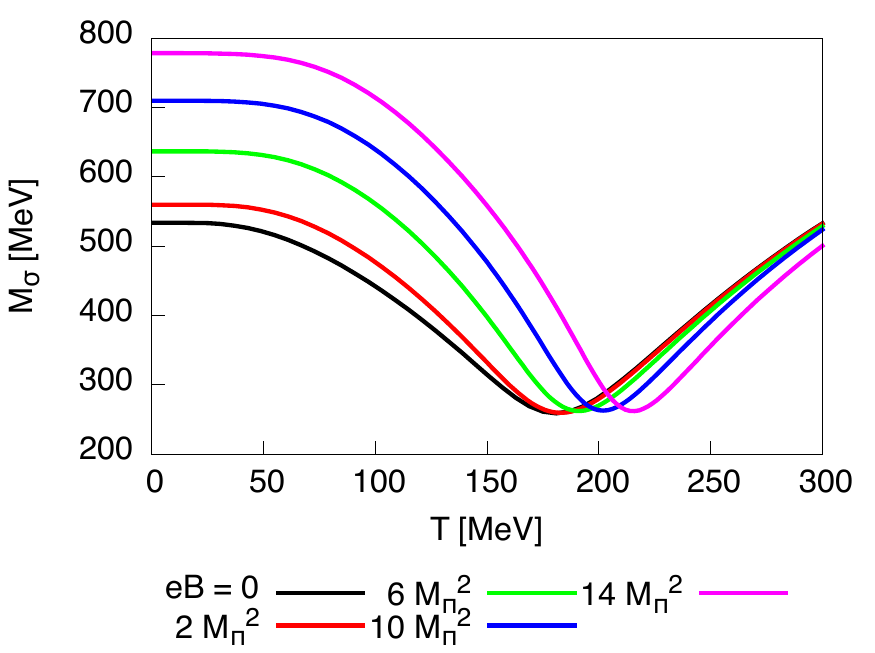}
     \quad 
     \includegraphics[width=0.48\columnwidth]{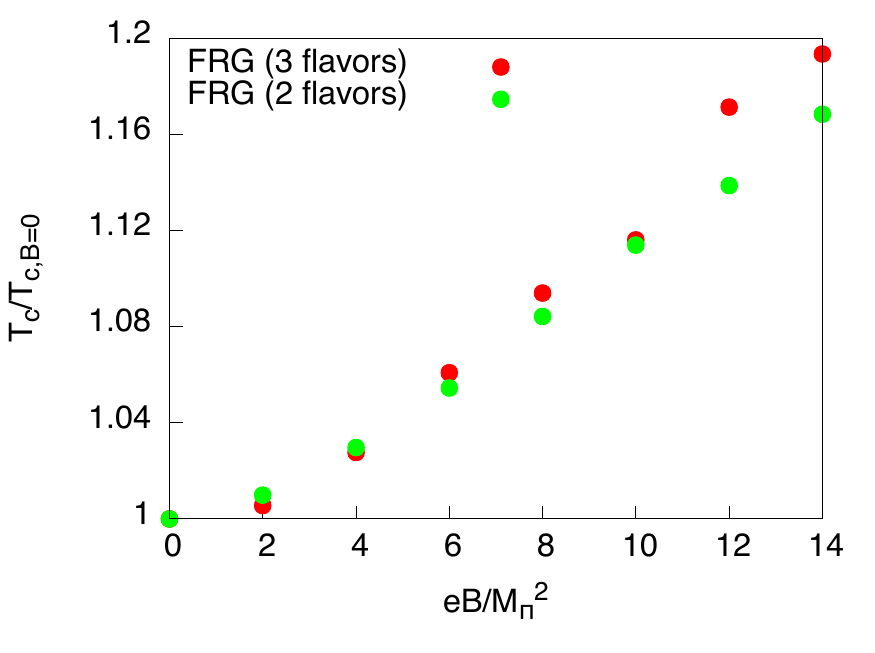}
   \end{center} 
   \vspace{-1.1\baselineskip}
   \caption{
     \label{fig:tc_in_magnetic_field}
     \textbf{Left:} The sigma meson mass from the three-flavor FRG.   
     \textbf{Right:} Chiral pseudo-critical temperature, determined from the minimum of the sigma mass, from 
     the two- and three-flavor FRG with a varying external magnetic field.
   }
   \end{figure}
  
  In this section we show numerical results for the particle masses, pressure, 
  pion and kaon decay constants and the chiral pseudo-critical temperature
  under an external magnetic field.  
  
  In the left panel of Figure \ref{fig:tc_in_magnetic_field}, we show the sigma meson mass $M_\sigma$ 
  obtained from the three-flavor FRG as a function of $T$ across the chiral crossover. At low 
  temperatures, $M_\sigma$ increases rapidly with $B$. The temperature at which 
  $M_\sigma$ reaches the bottom increases from around $180$ MeV at $B=0$ to higher values 
  for stronger $B$.  In the right panel of Figure \ref{fig:tc_in_magnetic_field}, we plot 
  the chiral pseudo-critical temperature $T_c$ defined as the temperature at which 
  $M_\sigma$ hits the bottom. It is found that $T_c$ increases with $B$ in both 
  the two- and three-flavor FRG. Although this feature is commonly seen in almost all chiral effective 
  models, it is at variance with lattice calculations at the physical point \cite{Bali:2011qj} 
  which reports a \emph{decrease} of $T_c$ at least for $eB<1$ GeV$^2$\,. 
  We conclude that the undesired behavior of the model 
  cannot be cured by inclusion of fluctuations of strange quark and $\SU(3)$ mesons. We believe that 
  $T_c$ rising with $B$ is not an artifact of LPA, because the inclusion of the wave function renormalization 
  for mesons leads to an even steeper increase of $T_c$ in the one-flavor QM model \cite{Kamikado:2013pya}. 

   \begin{figure}[!t]
   \begin{center}
    \mbox{
    \includegraphics[width=0.48\columnwidth]{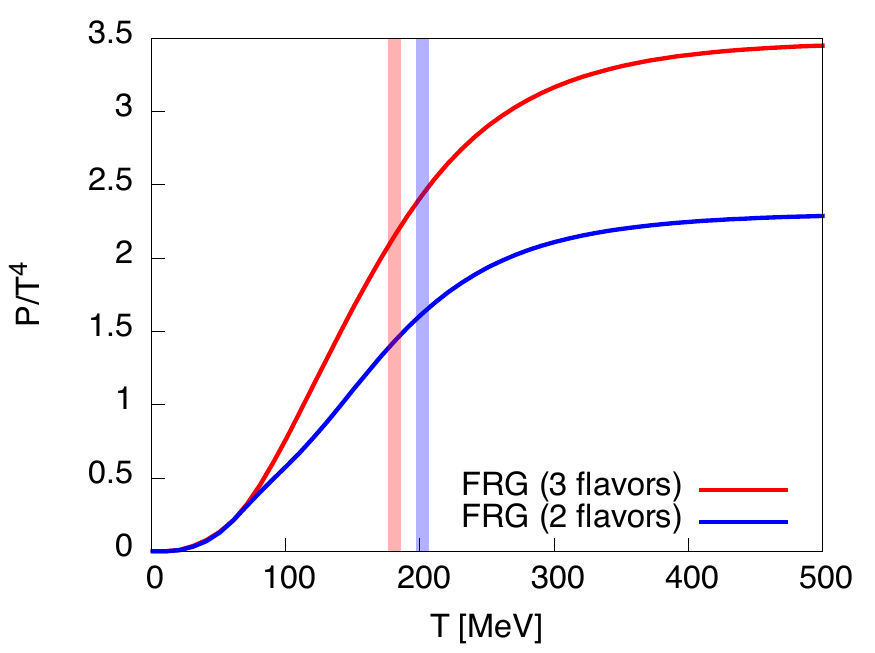}
    \quad
    \includegraphics[width=0.48\columnwidth]{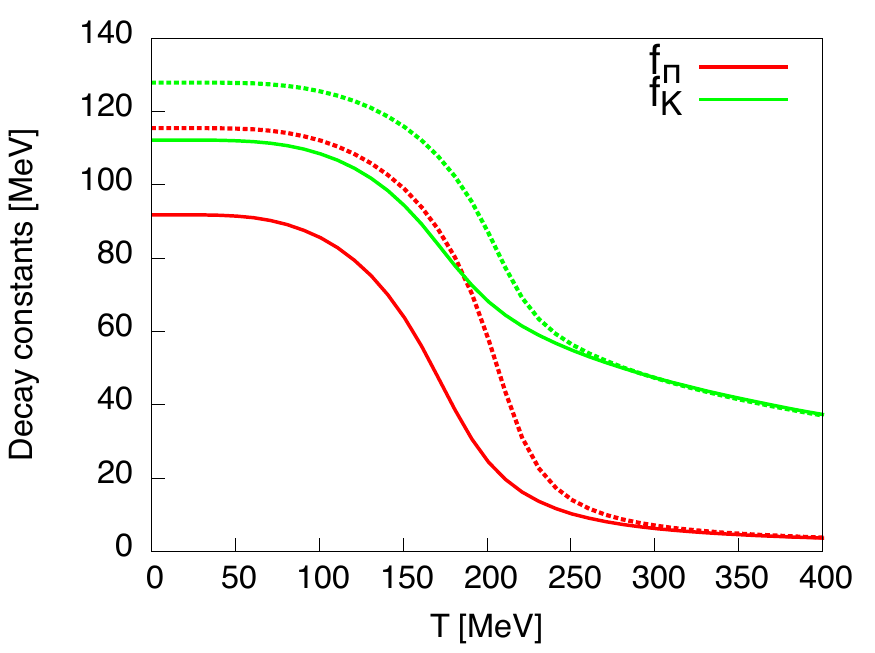}
    }
   \end{center} 
   \vspace{-\baselineskip}
   \caption{
     \label{fig:pressure_vs_t}
     \textbf{Left:}~Pressure from the two- and three-flavor FRG at $eB=0$. 
     Red and blue vertical bands indicate the pseudo-critical temperature in each theory.  
     \textbf{Right:}~The pion and kaon decay constant at finite temperature. 
     Solid and dashed lines correspond to $eB=0$ and $eB = 14
     m_{\pi}^2$, respectively. 
   } 
   \begin{center}
     \includegraphics[width=0.48\columnwidth]{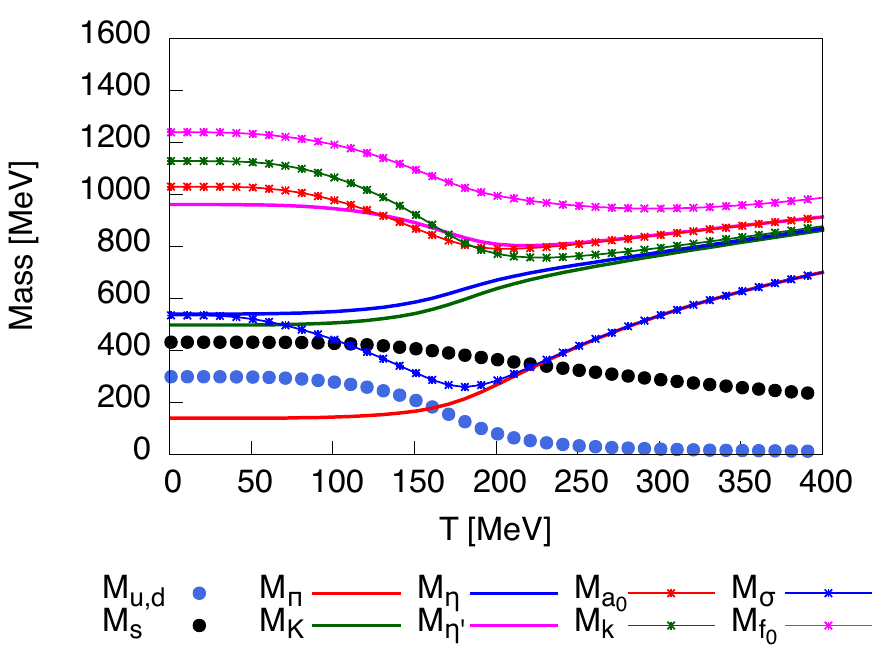}
     \quad 
     \includegraphics[width=0.48\columnwidth]{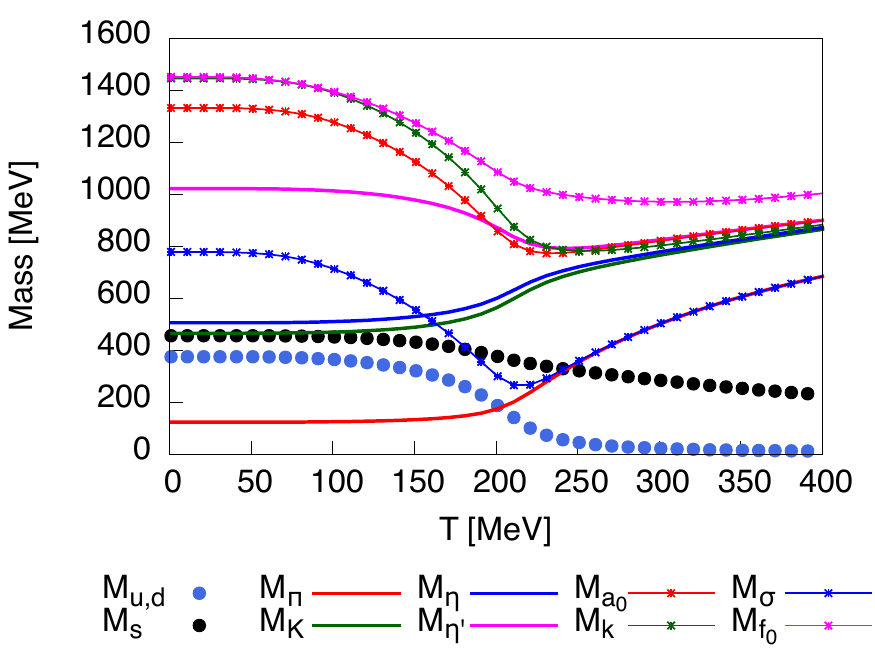}
     \put(-330,151){$eB=0$}
     \put(-118,151){$eB=14m_\pi^2$}
   \end{center}
   \vspace{-\baselineskip}
   \caption{
     \label{fig:particle_masses_su3}
     Masses of quarks (filled bullets), 
     pseudo-scalar mesons (solid lines) and scalar mesons 
     (solid lines with $\ast$) from the three-flavor FRG for
     $eB=0$ ({\bf Left}) and $eB=14m_\pi^2$ ({\bf Right}). 
     $M_k$ denotes the $\kappa$ meson mass. 
   }
   \end{figure}

  In the left panel of Figure \ref{fig:pressure_vs_t}, we plot the normalized pressure 
  $P/T^4$. For comparison, we display results from both the two- and three-flavor FRG 
  calculations. At high $T$, the pressure slowly converges to the Stephan-Boltzmann 
  limit of a free quark gas.%
  \footnote{%
    This behavior is due to the introduction of the residual part of the
    pressure ($P_r$) in \eqref{eq:P_residue}. We found that $P_r$ begins to matter at $T \gtrsim 200$ MeV.
  }
  At \mbox{$T\lesssim 70$ MeV} the two curves agree precisely, as expected from the fact that 
  pions dominate the pressure at low temperatures. In the right panel of Figure \ref{fig:pressure_vs_t}, 
  $f_\pi$ and $f_K$ are plotted for $eB=0$ and $eB=14m_\pi^2$. They increase with $B$ at all 
  temperatures, exhibiting the phenomenon of magnetic catalysis. We observe that $f_K$ decreases very slowly 
  compared to $f_\pi$\,, due to the large constituent mass of strange quark. 
     
  Figure \ref{fig:particle_masses_su3} shows the quark and meson masses 
  obtained from the three-flavor FRG calculation. The result for $eB=0$ (left panel) is consistent 
  with the preceding work \cite{Mitter:2013fxa}. At low temperatures, chiral symmetry is spontaneously 
  broken and quarks acquire large dynamical masses of order $300\sim400$ MeV. 
  At $T\gtrsim 180$ MeV, the light quark condensates begin to melt and sigma becomes degenerate 
  with pions, signalling the restoration of $\SU(2)\times \SU(2)$ chiral symmetry. On the other hand, 
  other mesons gain masses of order 1 GeV at high temperatures due to the large mass of strange quark. 
  These gross features persist in the presence of a magnetic field (right panel), 
  though the chiral crossover is shifted to a higher temperature ($\sim 220$ MeV for $eB=14m_\pi^2$). 
  Note that the anomaly strength is fixed in this work: if we implement the effective restoration of $\U_{\rm A}(1)$ 
  symmetry at high $T$, the meson spectrum would be changed qualitatively \cite{Pisarski:1983ms} 
  but it is beyond the scope of this work.

  \begin{figure}[tb]
   \begin{center}
    \includegraphics[width=0.62\columnwidth]{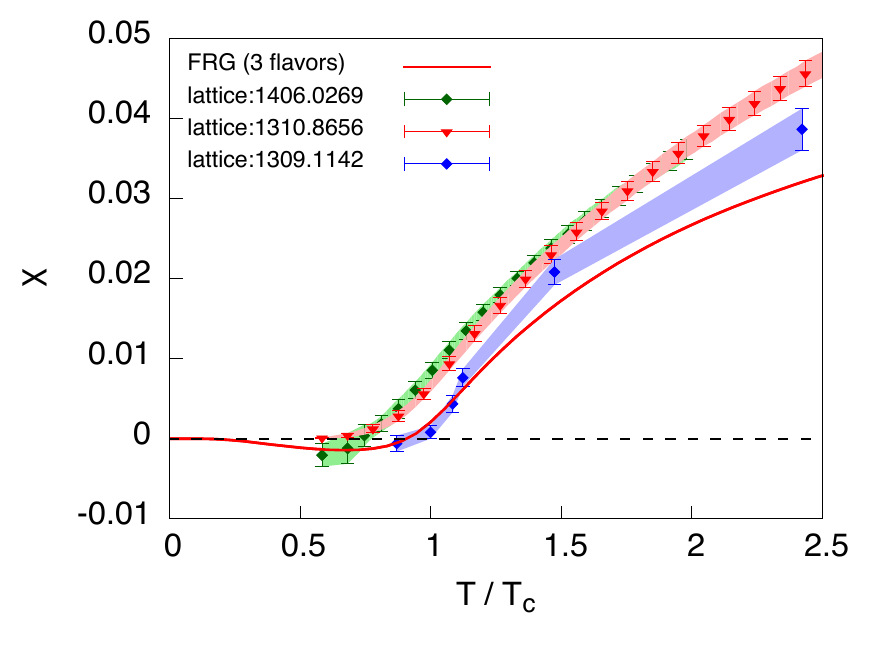}
   \end{center} 
   \vspace{-1.2\baselineskip}
   \caption{Magnetic susceptibility $\chi(T)$ from the three-flavor FRG calculation (solid line), in comparison with  
   the results of lattice QCD simulations by three different groups \cite{Levkova:2013qda,Bonati:2013vba,Bali:2014kia}.
   }  
   \label{fig:sus_vs_lat_normal_t}
  \end{figure}

  \subsubsection{Magnetic susceptibility}
  \label{sec:chi}
  
  This section is the main part of this paper. We compare $\chi(T)$ obtained from FRG 
  with the lattice QCD data. Although a perfect agreement with lattice cannot be expected due to 
  the schematic nature of the QM model, we believe such a comparison could help us develop an intuitive understanding for 
  gross features of QCD in a magnetic field. 
  
  In Figure \ref{fig:sus_vs_lat_normal_t}, $\chi(T)$ obtained with the $N_f=3$ FRG is plotted 
  together with the data from three independent lattice QCD simulations \cite{Levkova:2013qda,Bonati:2013vba,Bali:2014kia}.%
  \footnote{%
  In Figure \ref{fig:sus_vs_lat_normal_t}, the temperature is normalized by $T_c$ at $eB=0$: we used \mbox{$T_c=181$ MeV} for 
  the results of FRG and \mbox{$T_c=154$ MeV} \cite{Bazavov:2011nk} for all the lattice data. 
  This normalization makes a direct comparison of different methods easier.
  } 
  The figure shows a reasonable agreement between the FRG prediction and the lattice data over the entire 
  temperature range. In the high-$T$ QGP phase, quarks give a dominant paramagnetic contribution 
  to $\chi(T)$, which is well captured by our FRG calculation. Although the shape of the curve resembles the lattice data, 
  FRG seems to underestimate $\chi(T)$ at $T>T_c$ by about 30\%. Around $T_c$, FRG nicely agrees with the data 
  from \cite{Levkova:2013qda} but disagrees with those from \cite{Bonati:2013vba,Bali:2014kia}.     
  We do not understand 
  the origin of these discrepancies yet.%
  \footnote{The disagreement between \cite{Levkova:2013qda} and \cite{Bonati:2013vba,Bali:2014kia} 
  may be due to the larger quark mass in \cite{Levkova:2013qda}. We thank G.~Endr\H{o}di for a useful 
  comment on this point.}
  One possibility is that it is somehow related to the inability of the QM model to reproduce 
  inverse magnetic catalysis. Another possibility is that the paramagnetic contribution of spinful hadrons (e.g., $\rho^\pm$) 
  which are completely 
  ignored in the QM model has caused this discrepancy. Further investigation of this issue is left for future work. 

  Turning now to the hadronic phase below $T_c$\,, we observe that both FRG and the lattice data from \cite{Bali:2014kia} 
  yield negative values of $\chi(T)$, which are consistent with each other within error bars. This tendency could be explained 
  by diamagnetic contribution of light pions. Indeed, the region with $\chi(T)<0$ was not visible 
  in the mean-field calculation of the Polyakov linear sigma model \cite{Tawfik:2014hwa} which ignored meson fluctuations.  

  Overall, at a qualitative level, our three-flavor FRG correctly describes the transition of QCD 
  between diamagnetism and paramagnetism in the chiral crossover.  To the best of our knowledge, this is the first time such a 
  transition is demonstrated in a strongly interacting QCD-like theory. In the future it will be important to consolidate the diamagnetic 
  nature of QCD at low temperature by increasing lattice data points with better statistics.

  \begin{figure}[tb]
   \begin{center}
     \includegraphics[width=0.62\columnwidth]{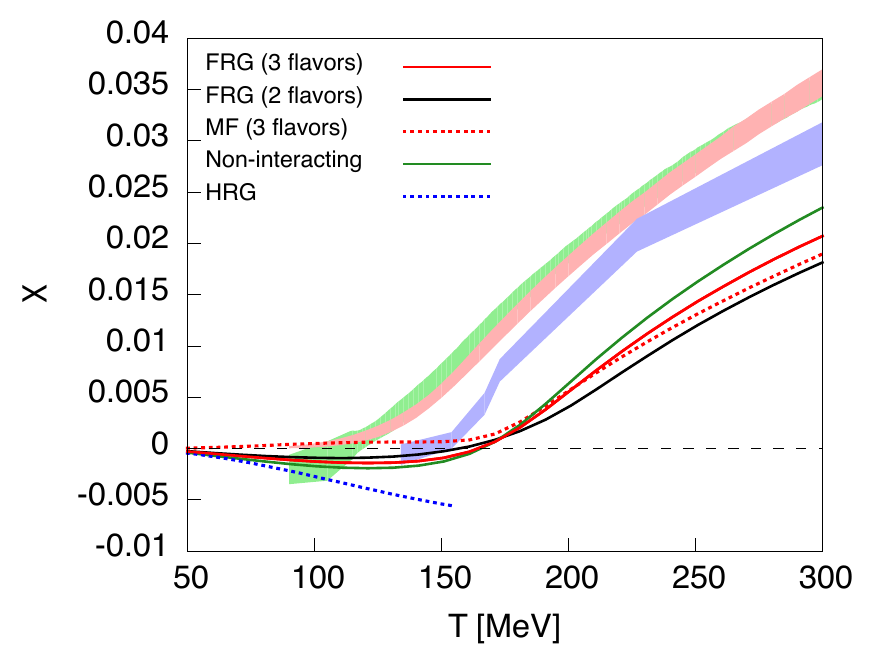}
   \end{center} 
   \vspace{-1.2\baselineskip}
   \caption{Magnetic susceptibilities obtained with different methods. 
   MF and HRG denote the mean-field approximation and the Hadron Resonance Gas model, 
   respectively. 
   Colored bands in the figure are the same as those in Figure \ref{fig:sus_vs_lat_normal_t} 
   and represent the results of lattice simulations \cite{Levkova:2013qda,Bonati:2013vba,Bali:2014kia}. 
   }  \label{fig:sus_vs_lat_shifted_t}
  \end{figure}

  In Figure \ref{fig:sus_vs_lat_shifted_t} we showcase a collection of results from various different methods.  
  Note that this time the horizontal axis is $T$ (without devision by $T_c$). Let us discuss characteristics 
  of each method one by one. 
  \begin{enumerate}
    \item 
    The magnetic susceptibility obtained from the Hadron Resonance Gas (HRG) model, calculated 
    with formulas in \cite[Appendix B]{Bali:2014kia}, is shown with a blue dashed line.  It agrees with the result of FRG 
    at $T\lesssim 70$ MeV because in this region the pressure in both calculations is dominated by 
    light pions, which behave diamagnetically \cite{Endrodi:2013cs,Bonati:2013vba,Bali:2014kia}. While 
    $\chi(T)$ from the HRG model is monotonically decreasing for all $T<T_c$\,, $\chi(T)$ from FRG stops decreasing 
    at $T\sim 130$ MeV: it is presumably because pions and kaons get heavier whereas quarks become lighter, as a result of 
    chiral restoration in the QM model. 
    \item 
    The results of two- and three-flavor FRG are shown in black and red solid lines. They share the same features ($\chi<0$ at 
    low $T$ and $>0$ at high $T$) and their difference is small over the entire temperature range. We conclude 
    that $\chi(T)$ is rather insensitive to the inclusion of strange quark and $\SU(3)$ mesons for $T\lesssim 300$ MeV.  
    \item 
    $\chi(T)$ from the mean-field approximation (MF) is plotted in a red dashed line. Not surprisingly, it is positive for all 
    $T$ and fails to capture the diamagnetism predicted by FRG and HRG at low $T$, because MF neglects 
    meson fluctuations altogether. This illustrates why it is imperative for the calculation of $\chi(T)$ to go beyond MF.  
    \item 
    As a hybrid of FRG and the HRG model, we considered a model in which quarks and meson are non-interacting 
    \emph{but have temperature-dependent masses.}  The magnetic susceptibility in such a model is derived 
    in appendix \ref{sc:vac_contr}. As inputs, we inserted the masses obtained with three-flavor FRG at $eB=0$ (shown in 
    Figure \ref{fig:particle_masses_su3}). 
    The resulting $\chi(T)$ is shown in Figure \ref{fig:sus_vs_lat_shifted_t} as a solid green line, 
    which mimics the results of FRG quite well at all temperatures. We conclude that the $T$-dependence of 
    the mass spectrum plays a vital role in determining the behavior of $\chi(T)$. 
  \end{enumerate}

\section{Conclusions and outlook}

 \label{sec:conclusion} In this work, we investigated thermodynamic properties 
 of a strongly interacting thermal medium under external 
 magnetic fields. We used the three-flavor quark-meson model with $\U_{\rm A}(1)$ anomaly, 
 which provides a physically consistent smooth 
 interpolation of a pion gas at low temperature and a free quark gas at high temperature. 
 In order to go beyond the mean-field approximation we utilized the 
 nonperturbative functional renormalization group (FRG) equation with the
 local-potential approximation which enables to incorporate of 
 fluctuations of charged and neutral mesons.

 Our results confirmed that within FRG the chiral pseudo-critical temperature increases with
 increasing external magnetic field. This is consistent with other chiral model approaches 
 to QCD with magnetic fields but does not concur with lattice QCD simulations at the physical quark masses. 
 We also calculated meson masses and decay constants as functions of $T$ and $eB$.  
 
 We then calculated the magnetic susceptibility $\chi(T)$ in this model by extracting the 
 $\calO\mkakko{(eB)^2}$ term in the pressure, which is the main result of this paper. 
 We found that $\chi<0$ in the hadron phase. This is due to the fact that, at low $T$, pressure is dominated
 by the light pion contributions which render the medium weakly diamagnetic. Our result is in quantitative agreement 
 with the lattice QCD data \cite{Bali:2014kia}. This diamagnetism is invisible in the mean-field approximation, because 
 meson fluctuations are not taken into account. On the other hand, at high $T$, we found $\chi>0$, whose values lie in the 
 same ballpark as the lattice QCD outputs \cite{Bonati:2013lca,Levkova:2013qda,Bonati:2013vba,Bali:2014kia}. This 
 behavior is caused by chiral symmetry restoration in the QGP phase, which makes quarks lighter and mesons heavier, 
 thereby letting the paramagnetic contribution of quarks dominate the magnetic susceptibility.   
 We observed that $\chi(T)$ crosses zero near the pseudo-critical temperature. While such a transition between 
 paramagnetism and diamagnetism has been known from a non-interacting MIT bag-type model \cite{Agasian:2008tb}, 
 this work has achieved for the first time a quantitatively reliable 
 demonstration of such a transition in a strongly interacting QCD-like model with spontaneous chiral symmetry breaking.  
 Our results indicate that bulk features of the response of QCD to \emph{weak} magnetic fields can be reproduced in the simple 
 quark-meson model, even though this model does not reproduce the inverse magnetic catalysis phenomenon in a \emph{strong} 
 magnetic field. Understanding of the remaining $30\%$ discrepancy of $\chi(T)$ at high temperature 
 between QCD and the model is an important open problem and we leave it for future work. 
 
 There are various directions to extend this work. From a technical point of view, one can systematically 
 improve the treatment of the quark-meson model (i) by introducing a scale-dependent 
 wave function renormalization and a flow of the Yukawa coupling $g$, 
 (ii) by allowing the effective potential to depend on another invariant 
 $\tr[(\Sigma \Sigma^\dagger)^3]$, (iii) by coupling quarks to the Polyakov loop background 
 to mimic confinement \cite{Skokov:2010wb,Herbst:2010rf,Skokov:2011ib}, and 
 (iv) by making the anomaly strength $c_A$ vary with temperature and/or magnetic field. 
 One can also use the framework of this paper to calculate the full equation of state, 
 magnetization, interaction measure, sound velocity, etc.~with possible inclusion of 
 a nonzero quark chemical potential. Such an extension will help to gain 
 better understanding of the physics of compact stars and heavy ion collisions. 
 
 \acknowledgments 
 The authors thank G.~S.~Bali, G.~Endr\H{o}di, T.~Hatsuda and Y.~Hidaka for
 valuable discussions. They are also grateful to useful comments from the
 participants of the sixth JICFuS seminar on 16 September 2014 
 and the workshop \emph{RIKEN iTHES workshop on Thermal
 Field Theory and its applications} on 3--5 September 2014, where part
 of this work has been presented.  KK was supported by the Special
 Postdoctoral Research Program of RIKEN. TK was supported by the RIKEN
 iTHES Project and JSPS KAKENHI Grants Number 25887014.

 \appendix
 \section{\boldmath Summary of the $N_f=2$ QM model}
 \label{sec:two-flavor-qm}
 The Lagrangian of the two-flavor QM model for $\bm{B}=0$ is given by
 \begin{equation}
   \label{eq:LagQM2}
   {\cal L} = \sum_{i=1}^{N_c}\bar{\psi}_i \big[
     \slashed{\partial} + g(\sigma+i \gamma_5 \vec{\pi} \cdot\vec{\tau}) 
   \big] \psi_i
   + \frac{1}{2}
   (\partial_{\mu} \sigma)^{2} + 
   \frac{1}{2}(\partial_{\mu} \vec{\pi})^{2}  + U (\rho) - h\sigma \,,
 \end{equation}
  where $\psi=(u,d)^T$ denotes the two-flavor quark field with three colors ($N_c=3$), and 
  $\rho \equiv \sigma^2 + \vec{\pi}^2$. The last term $-h\sigma$ represents the effect of a bare quark mass. 
  This model corresponds to the limit of infinitely strong $\U_{\rm A}(1)$ anomaly where $\eta$ and $a_0$ 
  decouple.
  
  To calculate the effective action of the model using the FRG equation, we employ LPA. Then 
  the Ansatz for the scale-dependent effective action with $\bm{B}\ne 0$ reads 
  \begin{equation}
   \begin{split}
    \label{eq:TB}
    \Gamma_k [\psi, \sigma, \pi] = & \int_0^{1/T}\!\!\! dx_4 \int d^3x ~
    \bigg\{ 
      \sum_{i=1}^{N_c} \bar\psi_i \big[ 
        \DD + g (\sigma + i\gamma_5 \vec\pi\cdot \vec\tau) 
    \big] \psi_i 
    \\
    & \qquad 
    + \frac{1}{2}(\partial_{\mu} \sigma)^2 + \mathcal{D}_\mu \pi^+ \mathcal{D}_\mu \pi^- + \frac{1}{2}(\partial_{\mu} \pi^0)^2 +
    U_k(\rho) -  h \sigma  \bigg\}\,,
   \end{split}
  \end{equation}
  where $\DD=\gamma_\mu(\partial_\mu-iQA_\mu)$ with $Q=\diag(\frac{2}{3}e, -\frac{1}{3}e)$, $\mathcal{D}_\mu=\partial_\mu+ieA_\mu$,  
  and $\pi^\pm=\frac{1}{\sqrt{2}}(\pi_1\pm i\pi_2)$. The effect of isospin symmetry breaking by the magnetic field on the 
  potential is neglected for simplicity.  
  With this Ansatz and the cutoff functions (\ref{eq:RB_3d}) and
  (\ref{eq:RF_3}), we can derive the flow equation for the effective
  potential (\ref{eq:flow_equation_for_potential}), in which the sum must now be restricted to 
  quarks and mesons of the two-flavor QM model.  The scale- and field-dependent energies 
  that enter the flow equation are defined as
  \begin{subequations}
   \begin{align}
    E_{u,d}(k)&= \sqrt{k^2 + g^2 \rho} \,, \\
    E_{\sigma}(k)&= \sqrt{k^2 + 2U_k' + 4\rho U_k''} \,, \\
    E_{\pi}(k)&=\sqrt{k^2 + 2U_k'}  \,,
   \end{align}
  \end{subequations}
  with $U_k' \equiv \partial_{\rho} U_k$ and $U_k'' \equiv \partial^2_{\rho} U_k$. 
  For the initial condition of the effective potential, we use
  \begin{equation}
   U_{k=\Lambda}(\rho) = a^{(1)}_\Lambda \rho + \frac{a^{(2)}_\Lambda}{2} \rho^2 \,.
  \end{equation}
  The two-flavor QM model has four free parameters besides the UV cutoff scale $\Lambda$. 
  We adjusted these parameters to reproduce the constituent quark mass, the sigma and pion masses 
  and the pion decay constant $f_{\pi} = \langle \sigma \rangle$, as summarized in Table~\ref{tab:parameter_2flavor}.
  \begin{table*}[h]
   \centering
   \begin{tabular}{|c || cccc | cccc|}
    \hline
    & $g$
    & $a_\Lambda^{(1)^{\mathstrut}}/\Lambda^2$ 
    & $a_\Lambda^{(2)}$ 
    & $h/\Lambda^3$  
    & $f_\pi$ & $m_\pi$ & $m_\sigma$ & $M_{u,d}$
    \\ \hline \hline 
    FRG ($N_f=2$) & 3.2 & $0.152$ & $5.0$ &$1.82 \times 10^{-3}$ & 
    $92.5$ & $140.3$ & $611.3$ & $296.1$
    \\ \hline   
   \end{tabular}
   \caption{Initial conditions at $k=\Lambda$ and resulting observables at $k=0$ 
   (in units of MeV) at $eB=T=0$ for $N_f=2$ FRG. 
   We used $\Lambda=1$ GeV. 
   }
   \label{tab:parameter_2flavor}
  \end{table*}

  \section{\boldmath Mass spectrum in the $N_f=3$ QM model}
  \label{sec:mass-eigenv-three}
  
  In this appendix, we explicitly present the scale-dependent screening masses $M_i$ of mesons in the
  three-flavor QM model, which enter the flow equation \eqref{eq:flow_equation_for_potential} 
  via the scale-dependent energy $E_b(k)$. A detailed derivation of the mass matrix 
  can be found in \cite{Mitter:2013fxa} and will not be reproduced here. 
  
  Let us switch from the strange-nonstrange basis \eqref{eq:sigmaxy} to a new coordinate defined by
  \begin{equation}
    X \equiv \sigma_x^2  \qquad \text{and} \qquad 
    Y \equiv 2\sigma_y^2 - \sigma_x^2 \;.
  \end{equation}
  The order parameter $X$ parametrises the magnitude of chiral symmetry breaking 
  in the light quark sector, whereas $Y$ represents the $\SU(3)$ flavor symmetry breaking due
  to the strange quark mass. 
  In terms of the new coordinate, the invariants may be expressed as
  \begin{equation}
    \rho_1 = \frac{3X + Y}{4}  \qquad 
    \text{and} \qquad  
    \rho_2 = \frac{Y^2}{24}\,.
    \label{eq:transform_x_to_rho}
  \end{equation}

  The field-dependent squared masses $\{ M_i^2 \}$ of $9+9=18$ mesons are obtained as 
  eigenvalues of the $18 \times 18$ matrix $\Omega_k$ defined by 
  \begin{equation}
    (\Omega_k)_{ij} \equiv \partial_{i} \partial_{j} \big[ U_k(\rho_1,\rho_2) -
    c_A \xi \big] \,,  
    \quad \qquad i, j=1,2,\dots,18 \hspace{-30pt}
  \end{equation}
  where the derivatives are taken with respect to the meson basis 
  $(\sigma_x,\sigma_1,\dots,\sigma_7, \sigma_y,\pi_0,\dots,\pi_8)$. 
  We note that the explicit symmetry breaking terms $\propto h_{x,y}$ 
  are linear in meson fields and hence do not contribute to the matrix elements of $\Omega_k$, 
  even though they affect the location of the minimum of the effective potential. 
    
  With the chain rule, it is tedious but straightforward to calculate the mass eigenvalues. 
  Using the notation
  \begin{align}
    U_k^{(i,j)}(\rho_1,\rho_2) & \equiv \bigg(\frac{\der}{\der {\rho_1}}\bigg)^i 
    \bigg(\frac{\der}{\der {\rho_2}}\bigg)^j U_k(\rho_1,\rho_2)\,, 
    \label{eq:Ubibun}
  \end{align}
  we now write down the mass eigenvalues of all mesons in terms of $X$ and $Y$: 
  \begin{itemize}
  \item {\bf Scalar mesons} 
  \begin{subequations}
   \begin{align}
    M^2_{a_0}&= U_k^{(1,0)}+\frac{6X-Y}{6} U_k^{(0,1)} + \frac{\sqrt{X+Y}}{2}c_A \,,
    \\
    M^2_{\kappa}&= U_k^{(1,0)} + \frac{3X+2Y + 3\sqrt{X}\sqrt{X+Y}}{6}U_k^{(0,1)} +
    \frac{\sqrt{X}}{2}c_A \,,
    \\
    M^2_{\sigma} &=\frac{A_{\sigma} + B_{\sigma} + \sqrt{(A_{\sigma} -
    B_{\sigma})^2 + 4 C_{\sigma}^2 }}{2} \,,
    \\
    M^2_{f_0} &=\frac{A_{\sigma} + B_{\sigma} - \sqrt{(A_{\sigma} -
    B_{\sigma})^2 + 4 C_{\sigma}^2 }}{2} \,,
   \end{align}
  \end{subequations}
  with the definitions 
  \begin{subequations}
   \begin{align}
    A_{\sigma} & \equiv U_k^{(1,0)}  + X U_k^{(2,0)}     + \frac{XY^2}{36} U_k^{(0,2)}    +
    \frac{2X-Y}{6} U_k^{(0,1)}     - \frac{YX}{3} U_k^{(1,1)}   -
    \frac{\sqrt{X+Y}}{2}c_A \,, 
    \\ 
    B_{\sigma} & \equiv U_k^{(1,0)}  + \frac{X+Y}{2}U_k^{(2,0)}    +
    \frac{(X+Y)Y^2}{18} U_k^{(0,2)}    + \frac{2X+3Y}{3} U_k^{(0,1)}     +
    \frac{X(X+Y)}{3} U_k^{(1,1)} \,,
    \\ 
    C_{\sigma} & \equiv \frac{\sqrt{X}\sqrt{X+Y}}{\sqrt{2}} U_k^{(2,0)}  
    - \frac{\sqrt{X} \sqrt{X+Y} Y^2}{18\sqrt{2}} U_k^{(0,2)}
    \notag 
    \\
    & \qquad - \frac{2\sqrt{X}\sqrt{X+Y}}{3 \sqrt{2}} U_k^{(0,1)}
    + \frac{\sqrt{X}\sqrt{X+Y}Y}{6\sqrt{2}} U_k^{(1,1)}   - \frac{\sqrt{X}}{\sqrt{2}}c_A \,.
   \end{align}
  \end{subequations}
  
  \item {\bf Pseudo-scalar mesons}
  \begin{subequations}
   \begin{align}
    M^2_{\pi}&= U_k^{(1,0)} -\frac{Y}{6}U_k^{(0,1)} - \frac{\sqrt{X+Y}}{2}c_A \,, 
    \\
    M^2_{K}&= U_k^{(1,0)} + \frac{3X+2Y-3\sqrt{X}\sqrt{X+Y}}{6}U_k^{(0,1)} - \frac{
    \sqrt{X}}{2}c_A \,, 
    \\
    M^2_{\eta} &=\frac{A_{\eta} + B_{\eta} + \sqrt{(A_{\eta} -
    B_{\eta})^2 + 4 C_{\eta}^2 }}{2} \,,
    \\
    M^2_{\eta'} &=\frac{A_{\eta} + B_{\eta} - \sqrt{(A_{\eta} -
    B_{\eta})^2 + 4 C_{\eta}^2 }}{2} \,,
   \end{align}
  \end{subequations}
  with the definitions 
  \begin{subequations}
   \begin{align}
    A_{\eta} & \equiv U_k^{(1,0)} + \frac{2\sqrt{X} + \sqrt{X+Y}}{3}c_A \,,
    \\
    B_{\eta} & \equiv U_k^{(1,0)} + \frac{Y}{6} U_k^{(0,1)} - \frac{4\sqrt{X}-\sqrt{X+Y}}{6}c_A \,,
    \\
    C_{\eta} & \equiv -\frac{Y}{3\sqrt{2}} U_k^{(0,1)} - 
    \frac{\sqrt{X} - \sqrt{X+Y}}{3\sqrt{2}}c_A \,. 
   \end{align}
  \end{subequations}
  \end{itemize}

  \section{Taylor method}
  \label{sec:taylor-method} 
  
  The flow equation for the effective potential can be solved numerically either by a grid method 
  or by the Taylor method. In this work we adopt the latter approach, in which the potential at a 
  scale $k$ is Taylor-expanded around the scale-dependent minimum. This works for any types of 
  flow equations and is very powerful except when a strong first-order phase transition occurs. 
  Since the chiral restoration is a smooth crossover in both two- and three-flavor QM model with 
  physical quark masses, one can safely rely on the Taylor method to solve the flow equation. 
  In this appendix, we derive flow equations for 
  the Taylor coefficients of the effective potential in the two- and three-flavor QM model.

  \subsection{Two flavors}
  First we consider the flow equation in the two-flavor QM model. In this case, the potential is a
  function of only one variable $\rho=\sigma^2+\vec{\pi}\,^2$ (recall \eqref{eq:LagQM2}). 
  Since the quark mass induces a condensate in $\sigma$-direction, we set $\vec\pi=\vec{0}$ so that 
  $\sigma=\sqrt{\rho}$. Now we expand the scale-dependent effective potential around the scale
  dependent minimum ($\rho_k$),
  \begin{equation}
   \begin{split}
    \bar{U}_k(\rho) & \equiv U_k(\rho) - h \sqrt{\rho}
    \\
    & \equiv \sum_{n=0}^{\infty} \frac{a_n}{n!}  (\rho-\rho_k)^n - h \sqrt{\rho}\,,
   \end{split}
  \end{equation}
  where the coefficients $\{a_n\}$ are functions of $k$. Differentiating this expression 
  with respect to $k$, we obtain an infinite family of flow equations for $\{a_n\}$,
  \begin{equation}
    \label{eq:anQM2flow}
    \dd_k a_n = \der_k U_k^{(n)}\Big|_{\rho = \rho_k} + a_{n+1} \dd_k \rho_k \,,
  \end{equation}
  with $U_k^{(n)} \equiv \der^{n} U_k/ \partial \rho^n$ and $\displaystyle \dd_k \equiv \frac{\dd}{\dd k}$. 
  On the RHS, $\der_k U_k^{(n)}$ can be obtained by differentiating the flow equation of $U_k$ 
  with $\rho$, whereas $\dd_k \rho_k$ may be deduced as follows: First, recall that 
  the scale-dependent minimum of the effective potential should satisfy the following
  minimum condition at any scale:%
  \footnote{This minimum condition is meaningful only if the condensate is non-vanishing ($\rho_k>0$). 
  Fortunately, in the present work, the quark mass effect $\propto - h\sqrt{\rho}$ always guarantees  
  a non-vanishing condensate.}
  \begin{equation}
    \partial_{\rho} \bar{U}_k\big|_{\rho = \rho_k} \overset{!}{=}0 \quad \Rightarrow \quad  a_1
    = \frac{h}{2 \sqrt{\rho_k}} \quad \text{for}~~^\forall k\;. 
    \label{eq:rhoflow}
  \end{equation}
  Combining \eqref{eq:rhoflow} with \eqref{eq:anQM2flow} for $n=1$, we obtain 
  the flow equation for the scale-dependent minimum 
  \begin{align}
    \dd_k \rho_k &=  -\frac{ \partial_k U^{(1)}_k\big|_{\rho = \rho_k}}{\frac{h}{4 \rho_k^{3/2}} + a_2} \,.
  \end{align}
  In this work, we carried out the Taylor expansion up to $n=6$.

  \subsection{Three flavors}
  \label{sc:taylor2+1}
  
  Next, we consider the flow equation in the three-flavor QM model. 
  Assuming that only $\sigma_0$ and $\sigma_8$ take nonzero values, we obtain   
  $\Sigma=\diag\big( \sigma_x/2,~\sigma_x/2,~\sigma_y/\sqrt{2} \big)$ with 
  $\sigma_{x,y}$ defined in \eqref{eq:sigmaxy}. Plugging this into the definition of 
  $\rho_{1,2}$, we find $\rho_1=(\sigma_x^2+\sigma_y^2)/2$ and 
  $\frac{1}{3}\rho_1^2+\rho_2=(\sigma_x^4+2\sigma_y^4)/8$. 
  It is then straightforward to solve them for $\sigma_x$ and $\sigma_y$. 
  
  What is new for $N_f=3$ compared to $N_f=2$ is that the total effective potential 
  in \eqref{eq:lpa_effective_action} is a function of \emph{two} variables, 
  \begin{equation}
   \begin{split}
    \bar{U}_k(\rho_1,\rho_2) & \equiv U_k(\rho_1,\rho_2) - h_x \sigma_x 
    - h_y \sigma_y - c_A \xi 
    \\
    & = U_k(\rho_1,\rho_2) + f(\rho_1,\rho_2)\,,
   \end{split}
  \end{equation} 
  with 
  \begin{equation}
    f(\rho_1,\rho_2) \equiv 
    - h_x \sqrt{\frac{4\rho_1-\sqrt{24 \rho_2}}{3}} 
    - h_y \sqrt{\frac{2\rho_1+\sqrt{24 \rho_2}}{3}} 
    - \frac{c_A}{2\sqrt{2}} \frac{4\rho_1-\sqrt{24\rho_2}}{3} 
    \sqrt{\frac{2\rho_1+\sqrt{24 \rho_2}}{3}}\;.
  \end{equation} 
   
   Then we have to generalize the Taylor method into two variables. The main
   idea is same as the two-flavor case: we expand $U_k$ around 
   the scale-dependent minimum $(\rho_{1,k},\rho_{2,k})$ of $\bar{U}_k$ as 
   \begin{equation}
     U_{k}(\rho_1,\rho_2) = \sum_{i,j=0}^{\infty} \frac{a_{i,j} }{i! j!}
     (\rho_1-\rho_{1,k})^{i}(\rho_2-\rho_{2,k})^{j}\,,
     \label{eq:Uexpand}
   \end{equation}
   where $\{a_{i,j}\}$ are $k$-dependent coefficients. 
   In this work, we take into account the coefficients up to third order of invariants, 
   i.e., $a_{i,j} \; (i,j \le 3)$. For convenience, let us define $\kkakko{ \mathcal{F}(\rho_1,\rho_2) }_k
   :=\mathcal{F}(\rho_{1,k},\rho_{2,k})$ for an arbitrary function $\mathcal{F}$ of 
   $\rho_1$ and $\rho_2$. 
    
    By differentiating both sides of \eqref{eq:Uexpand} with respect to $k$ 
    we obtain an infinite tower of flow equations for the coefficients: 
    \begin{equation}
      \dd_{k} a_{i,j} = \kkakko{\der_{k}U_k^{(i,j)}}_k + a_{i+1,j} \, \dd_{k}
      \rho_{1,k}  + a_{i,j+1} \, \dd_{k} \rho_{2,k} \,,
      \label{eq:flow_coef}
    \end{equation}
    with $U_k^{(i,j)}$ defined in \eqref{eq:Ubibun}. 
    On the RHS, $\der_k U_k^{(i,j)}$ can be obtained by differentiating 
    $\der_k U_k$ in \eqref{eq:flow_equation_for_potential}  
    with respect to $\rho_1$ and $\rho_2$, whereas $\dd_k \rho_{1,k}$ and 
    $\dd_k\rho_{2,k}$ are derived as follows: First, 
    the minimum condition of $\bar U_k$ at $(\rho_{1,k},\rho_{2,k})$ reads%
    \footnote{%
      It must be noted that these minimum conditions are meaningful 
      only if $\rho_{1,k}>0$ and $\rho_{2,k}>0$. This is guaranteed for nonzero 
      quark masses, i.e., $h_{x,y}>0$. 
    }   
    \begin{subequations}
     \begin{align}
      \kkakko{\der_{\rho_1} \bar{U}_k}_k \overset{!}{=} 0 & 
      \quad \Longleftrightarrow \quad 
      a_{1,0} = - \kkakko{\partial_{\rho_1} f}_k \,,
      \\
      \kkakko{\der_{\rho_2} \bar{U}_k}_k \overset{!}{=} 0 & 
      \quad \Longleftrightarrow \quad 
      a_{0,1} = - \kkakko{\partial_{\rho_2} f}_k \,. 
     \end{align}
    \end{subequations}
    By differentiating the above relations with respect to $k$ and using \eqref{eq:flow_coef}, we obtain
    \begin{subequations}
      \begin{align}
       a_{2,0} \dd_k \rho_{1,k}+ a_{1,1} \dd_k\rho_{2,k} + \kkakko{\der_k U_k^{(1,0)}}_k 
       &= - \kkakko{ \partial^2_{\rho_1} f}_k \dd_k \rho_{1,k}
       - \kkakko{ \partial_{\rho_1}\partial_{\rho_2} f}_k \dd_k \rho_{2,k} \,,
       \\
       a_{1,1} \dd_k \rho_{1,k}+ a_{0,2} \dd_k \rho_{2,k} + \kkakko{ \der_k U_k^{(0,1)}}_k 
       & = - \kkakko{ \partial_{\rho_1}\partial_{\rho_2} f}_k \dd_k {\rho}_{1,k} 
        - \kkakko{ \partial^2_{\rho_2} f}_k \dd_k \rho_{2,k} \,. 
      \end{align}
    \end{subequations}
    By solving these equations for $\dd_k \rho_{1,k}$ and 
    $\dd_k \rho_{2,k}$, we finally arrive at 
    \begin{subequations}
     \begin{align}
      \dd_k \rho_{1,k} &= 
      \frac{ 
        - \mkakko{ \kkakko{ \partial_{\rho_1}\partial_{\rho_2} f}_k + a_{1,1} }
        \kkakko{ \der_k U_k^{(0,1)} }_{k} 
        + \mkakko{ \kkakko{ \partial^2_{\rho_2} f}_k   + a_{0,2} }
        \kkakko{\der_k U_k^{(1,0)}}_k 
      }{ 
        \mkakko{ \kkakko{ \partial_{\rho_1}\partial_{\rho_2} f}_k + a_{1,1} }^2 -
        \mkakko{ \kkakko{ \partial^2_{\rho_1} f}_k +a_{2,0} }
        \mkakko{ \kkakko{ \partial^2_{\rho_2} f}_k + a_{0,2} }
      }  \,,
      \\
      \dd_k \rho_{2,k} &= 
      \frac{
        - \mkakko{ \kkakko{ \partial_{\rho_1}\partial_{\rho_2} f}_k+a_{1,1} } \kkakko{\der_k U_k^{(1,0)}}_k +
        \mkakko{\kkakko{ \partial^2_{\rho_1} f}_k + a_{2,0} } \kkakko{\der_k U_k^{(0,1)}}_k 
      }
      {
        \mkakko{ \kkakko{ \partial_{\rho_1}\partial_{\rho_2} f}_k + a_{1,1} }^2 -
        \mkakko{ \kkakko{ \partial^2_{\rho_1} f}_k  + a_{2,0} } \mkakko{\kkakko{ \partial^2_{\rho_2} f}_k +a_{0,2}}   
      } \,.
     \end{align}
     \label{eq:flow_order_param}
    \end{subequations}

 \section{Magnetic susceptibility of a non-interacting quark-meson gas}
 \label{sc:vac_contr}
 
The goal of this appendix is to derive the magnetic susceptibility of a non-interacting gas 
of quarks and mesons with \emph{temperature-dependent masses}. Our new result is 
$\chi(T) = \chi_q(T) + \chi_m(T)$\,, with the contribution from quarks 
\begin{align}
  \scalebox{0.95}{$\displaystyle 
  \chi_q(T) = \frac{N_c}{6\pi^2}  
  \sum_{f} \mkakko{\frac{e_f}{e}}^2 \bigg\{
  2 \int_0^\infty \!\!\! \frac{dx}{\sqrt{x^2+1}}
  \frac{1}{\exp\Big(\frac{m_f(T)}{T}\sqrt{x^2+1}\,\Big)+1
  }
  - \log \mkakko{\frac{m_f(0)}{m_f(T)} }
  \bigg\}
  $}\,,
  \label{eq:q_sus}
\end{align}
and from charged mesons
\begin{align}
  \chi_m(T) & = -\frac{1}{48 \pi^2} \sum_{b} \bigg\{ 2 \int_0^\infty\!\!\!
  \frac{dx}{\sqrt{x^2+1}}\frac{1}{\exp\Big(\frac{m_b(T)}{T}\sqrt{x^2+1}\,\Big)-1}
  + \log \mkakko{\frac{m_b(0)}{m_b(T)}} \bigg\}  \,,
  \label{eq:m_sus}
\end{align}
with $T$-dependent masses $m_f(T)$ and $m_b(T)$, respectively. 
In \eqref{eq:m_sus}, the index $b$ runs over $\pi^+, \pi^-, K^+, K^-, a_0^+, a_0^-, \kappa^+$ and $\kappa^-$. 
The first terms inside brackets in (\ref{eq:q_sus}) and (\ref{eq:m_sus}) are thermal contributions 
that have already been computed in \cite{Agasian:2008tb}.%
\footnote{There is a typo in the first equation of \cite[Eq.(30)]{Agasian:2008tb}: 
$12\pi^2$ in the denominator should read $24\pi^2$. We checked this with a numerical software.} 
The second terms in (\ref{eq:q_sus}) and (\ref{eq:m_sus}) are new vacuum 
corrections, which we are going to work out below. 

Let us begin with the vacuum energy density for a particle with spin $s$, charge $q$ 
and mass $m$ in a magnetic field:
 \begin{align}
   f^{\rm vac}(m) & \equiv \mp \frac{1}{2}\sum_{n=0}^{\infty}\sum_{s_z}\frac{|qB|}{2\pi}\int 
   \frac{dp_z}{2\pi} \sqrt{p_z^2+m^2+2|qB|(n+1/2-s_z)}\,,
   \label{eq:fvacfir}
 \end{align}
 where the upper sign corresponds to fermions and the lower to bosons. If $m$ is a constant 
 mass, then $f^{\rm vac}(m)$ is independent of $T$ and does not contribute to the subtracted pressure, 
 \eqref{eq:normalized_pressure}. However this is not the case if $m$ depends on $T$,  
 as we will shortly see.   

 As mentioned in section \ref{sec:obs}, the vacuum energy density contains a $B$-dependent divergence 
 that has to be regularized. We follow the renormalization scheme in \cite{Endrodi:2013cs}. 
 With dimensional regularization and using dimensionless variables $a\equiv \frac{1}{2}-s_z$  
 and $x\equiv \frac{m^2}{2|qB|}$\,, we obtain (cf.~\cite[Eq.(3.13)]{Endrodi:2013cs})
 \begin{align}
  & f^{\rm vac}(m)\big|_{B\ne 0} - f^{\rm vac}(m)\big|_{B=0} 
  \notag
  \\
  & = \scalebox{0.85}{$\displaystyle \pm \frac{(qB)^2}{8\pi^2}\sum_a \bigg[ 
  \mkakko{\frac2{\epsilon}-\gamma-\log\mkakko{\frac{2|qB|}{4\pi \mu^2}}+1}\mkakko{\frac{1}{12}-\frac{a}{2}+\frac{a^2}{2}}
  - \zeta'(-1,x+a) - \frac{x^2}{4}+\frac{x^2}{2}\log x
  \bigg]$ } 
  \label{eq:D4}
  \\
  & =:\Delta f^{\rm vac}(m)\,, \notag
 \end{align}
 where $\mu$ is an arbitrary scale introduced for a dimensional reason. 
 The renormalization prescription in \cite{Endrodi:2013cs} is to ensure that the quadratic term 
 in the renormalized vacuum energy density only consists of a pure magnetic-field contribution, namely
 \begin{align}
   \Big(\Delta f^{\rm vac}(m)+\frac{B^2}{2}\Big)\bigg|_{m=m_\star} & \overset{!}{=} \frac{B_r^2}{2}+\calO(B^4)\,, 
   \label{eq:fvacr}
 \end{align}
 where we have defined 
 \begin{align}
   B^2 = Z_q(m_\star) B_r^2\,, \quad 
   q^2 = \frac{1}{Z_q(m_\star)}q_r^2 \,, \quad \text{and} \quad 
   q^2B^2 = q_r^2 B_r^2\,,
 \end{align} 
 with the wave-function renormalization factor 
 \begin{align}
    Z_q(m_\star) & \equiv  1 \mp \frac{q_r^2}{8\pi^2} 
    \ckakko{\frac{2}{\epsilon}-\gamma-\log\mkakko{\frac{m_\star^2}{4\pi\mu^2}}}
    \sum_a 
    \mkakko{\frac{1}{6}-a+a^2} . 
 \end{align}
 Here $m_\star$ is a fixed mass scale. 
 
 Now we are prepared to consider a $T$-dependent mass: $m\to m(T)$. 
 To fulfill the condition \eqref{eq:fvacr} at $T=0$, we must set $m_\star=m(0)$\,. 
 Then, with $x=\frac{m(T)^2}{2|qB|}$ and using \eqref{eq:D4}, we obtain 
 \begin{align*}
  & \Delta f^{\rm vac}(m(T)) + \frac{B^2}{2}
  \\
  =~& \Delta f^{\rm vac}(m(T)) + \frac{B_r^2}{2}Z_q(m(0)) 
  \\
  =~&   
  \scalebox{0.96}{$\displaystyle \pm \frac{(qB)^2}{8\pi^2}\sum_a \bigg[ 
  \mkakko{\frac2{\epsilon}-\gamma-\log\mkakko{\frac{2|qB|}{4\pi \mu^2}}+1}\mkakko{\frac{1}{12}-\frac{a}{2}+\frac{a^2}{2}}
  - \zeta'(-1,x+a) - \frac{x^2}{4}+\frac{x^2}{2}\log x 
  \bigg] 
  $}
  \\
  & \quad + \frac{B_r^2}{2}\ckakko{1 \mp \frac{q_r^2}{8\pi^2} 
  \ckakko{\frac{2}{\epsilon}-\gamma-\log\mkakko{\frac{m(0)^2}{4\pi\mu^2}}}
  \sum_a \mkakko{\frac{1}{6}-a+a^2}}
  \\
  =~ & \frac{B_r^2}{2}\pm \frac{(qB)^2}{8\pi^2}\sum_a \bigg[ 
  \mkakko{\log\mkakko{\frac{m(0)^2}{2|qB|}}+1}\mkakko{\frac{1}{12}-\frac{a}{2}+\frac{a^2}{2}}
  - \zeta'(-1,x+a) - \frac{x^2}{4}+\frac{x^2}{2}\log x
  \bigg] \,.
 \end{align*}
 In the weak field limit $x\gg 1$, we use the expansion 
 \begin{align}
  \zeta'(-1,x+a) & =  \frac{x^2}{2}\log x - \frac{x^2}{4} +
  \mkakko{\frac{1}{12} - \frac{a}{2} +\frac{a^2}{2} }\mkakko{\log x+1} +
  \mkakko{a-\frac{1}{2}}x \log x 
  + \calO(x^{-2}) 
  \notag
 \end{align}
 and $\sum_a (a-1/2)=0$ to obtain
 \begin{align}
  & \Delta f^{\rm vac}(m(T)) + \frac{B^2}{2} 
  \notag
  \\
  = ~ & 
  \frac{B_r^2}{2}\pm \frac{(qB)^2}{8\pi^2} 
  \ckakko{\log\mkakko{\frac{m(0)^2}{2|qB|}}-\log\mkakko{\frac{m(T)^2}{2|qB|}}}
  \sum_a\mkakko{\frac{1}{12}-\frac{a}{2}+\frac{a^2}{2}} 
  + \calO(B^4)
  \\
  = ~ & \frac{B_r^2}{2}\pm \frac{(qB)^2}{8\pi^2}
  \log \mkakko{\frac{m(0)}{m(T)}} \sum_a\mkakko{ \frac{1}{6} - a + a^2 } 
  + \calO(B^4) \,. 
  \label{eq:fvacfin}
 \end{align}

 In the following we consider quarks and mesons separately. 
 
  \subsection{Quarks}

  For fermions with $s=1/2$, the sum in \eqref{eq:fvacfin} over $a=0$ and $1$ is trivial and yields
  \begin{align}
    \Delta f^{\rm vac}(m(T)) + \frac{B^2}{2}   & = \frac{B_r^2}{2} + \frac{(qB)^2}{24\pi^2}
    \log \mkakko{\frac{m(0)}{m(T)}} 
    + \calO(B^4) \,. 
    \label{eq:fvacqq}
  \end{align}
  The total vacuum energy density of all quarks at finite $T$ is given by
  \begin{align}
    V_{q}^{\rm vac}(T,B) & \equiv - N_c \sum_{f} \sum_{n=0}^{\infty}\sum_{s_z} \frac{|e_f B|}{2\pi}
    \int \frac{dp_z}{2\pi}\sqrt{p_z^2+m_f(T)^2+(2n+1-2s_z)|e_fB|}\,. 
  \end{align}
  Comparing $V_{q}^{\rm vac}(T,B)$ with \eqref{eq:fvacfir} and using \eqref{eq:fvacqq}, we find for the total free energy
  \begin{align}
   \frac{B^2}{2} + V_{q}^{\rm vac}(T,B) - V_{q}^{\rm vac}(T,0) 
   & =  \frac{B^2}{2} + 2 N_c \sum_{f} \Delta f^{\rm vac}(m_f(T)) \Big|_{q=e_f}
   \\
   & = \frac{B_r^2}{2} + 2 N_c \sum_{f} \frac{(e_f B)^2}{24\pi^2} \log \mkakko{\frac{m_f(0)}{m_f(T)}} 
   + \calO(B^4) \,. 
  \end{align}
  From this we obtain the vacuum correction to the magnetic susceptibility of quarks:
  \begin{align}
    \chi^{\rm vac}_q(T) & = - \frac{N_c}{6\pi^2} \sum_f \mkakko{\frac{e_f}{e}}^2 \log \mkakko{\frac{m_f(0)}{m_f(T)}} \,, 
  \end{align}
  which is nothing but the second term in \eqref{eq:q_sus}. 
  
  We note in passing that, in the high-temperature limit $\big(T\gg m_f(T)\big)$, the asymptotic behavior 
  of \eqref{eq:q_sus} becomes
  \begin{align}
   \chi_q(T) & \simeq \frac{N_c}{6\pi^2} \sum_f \mkakko{\frac{e_f}{e}}^2 
   \ckakko{
   \log \mkakko{\frac{T}{m_f(T)}} - \log \mkakko{\frac{m_f(0)}{m_f(T)}} 
   }
   \label{eq:chiqasym}
   \\
   & = \frac{N_c}{6\pi^2} \sum_f \mkakko{\frac{e_f}{e}}^2 \log \mkakko{\frac{T}{m_f(0)}} 
   \\
   & = 2\beta_1^{\rm QED} \log \mkakko{\frac{T}{m_f(0)}} \,, 
   \label{eq:chiqasym2}
  \end{align}
  where $\beta_1$ is the first coefficient of the QED beta function \cite{Elmfors:1993wj,Elmfors:1993bm,Endrodi:2013cs}. 
  Interestingly, the IR divergence in the chiral limit $(m_f(T)\to 0)$ neatly cancels out between the two 
  terms in \eqref{eq:chiqasym}! The final result \eqref{eq:chiqasym2} is well-defined and 
  finite, if $m_f(0)>0$ is dynamically generated. The necessity
  of a nonperturbative scale in the perturbative expression of magnetic
  susceptibility has been emphasized in \cite{Bali:2014kia} in the
  context of lattice QCD, while here we have extended their arguments to
  the case of a chiral effective model. It should be point out though, that the nonperturbative 
  scale for $\chi(T)$ extracted from lattice QCD is $\Lambda_{\rm H}=120$ MeV \cite{Bali:2014kia}, 
  which is smaller than $m_f(0)\sim 300$ MeV by a factor of 2.5\,.

  \subsection{Mesons}
  
  For spinless bosons with charge $q=e$, \eqref{eq:fvacfin} becomes 
  \begin{align}
    \Delta f^{\rm vac}(m(T)) + \frac{B^2}{2} 
    & = \frac{B_r^2}{2} + \frac{(eB)^2}{96\pi^2}\log \mkakko{\frac{m(0)}{m(T)}} 
    + \calO(B^4) \,. 
    \label{eq:Deltafm}
  \end{align}
  The total vacuum energy density of scalar and pseudo-scalar mesons at finite $T$ is given by
  \begin{align}
    V_{m}^{\rm vac}(T,B) & \equiv \frac{1}{2}\sum_{b} \sum_{n=0}^{\infty} \frac{|eB|}{2\pi}
    \int \frac{dp_z}{2\pi}\sqrt{p_z^2+m_b(T)^2+(2n+1)|eB|}\,, 
  \end{align}
  where the index $b$ runs over $\pi^+, \pi^-, K^+, K^-, a_0^+, a_0^-, \kappa^+$ and $\kappa^-$. 
  Comparing $V_{m}^{\rm vac}(T,B)$ with \eqref{eq:fvacfir} and using \eqref{eq:Deltafm}, 
  we find 
  \begin{align}
    \frac{B^2}{2} + V_{m}^{\rm vac}(T,B) - V_{m}^{\rm vac}(T,0)
    & = \frac{B^2}{2} + \sum_{b} \Delta f^{\rm vac}(m_b(T)) 
    \\
    & = \frac{B_r^2}{2} + \frac{(eB)^2}{96\pi^2} \sum_{b}
    \log \mkakko{\frac{m_b(0)}{m_b(T)}} + \calO(B^4)\,.
  \end{align}
  From this we obtain the vacuum correction to the magnetic susceptibility of mesons:
  \begin{align}
    \chi^{\rm vac}_m(T) & = - \frac{1}{48\pi^2} \sum_b \log \mkakko{\frac{m_b(0)}{m_b(T)}} \,, 
  \end{align}
  which is nothing but the second term in \eqref{eq:m_sus}.

  \bibliography{ref_frg_2}
  \bibliographystyle{JHEP}
 \end{document}